# Theory on the Dynamics of Oscillatory Loops in the Transcription Factor Networks

*Rajamanickam Murugan*

Department of Biotechnology, Indian Institute of Technology Madras
Chennai, 600036 India. Email: rmurugan@gmail.com

## Abstract

We develop a detailed theoretical framework for various types of transcription factor gene oscillators. We further demonstrate that one can build genetic-oscillators which are tunable and robust against perturbations in the critical control parameters by coupling two or more independent Goodwin-Griffith oscillators through either -OR- or -AND- type logic. Most of the coupled oscillators constructed in the literature so far seem to be of -OR- type. When there are transient perturbations in one of the -OR- type coupled-oscillators, then the overall period of the system remains constant (period-buffering) whereas in case of -AND- type coupling the overall period of the system moves towards the perturbed oscillator. Though there is a period-buffering, the amplitudes of oscillators coupled through -OR- type logic are more sensitive to perturbations in the parameters associated with the promoter state dynamics than -AND- type. Further analysis shows that the period of -AND- type coupled dual-feedback oscillators can be tuned without conceding on the amplitudes. Using these results we derive the basic design principles governing the robust and tunable synthetic gene oscillators without compromising on their amplitudes.

## Key words

Transcription factors, genetic oscillators, Hill coefficient, circadian rhythms, promoter state fluctuations

## Author Summary

Genetic oscillators drive various developmental as well as mitotic cell-cycle dynamics and circadian-rhythms associated with the intracellular concentration of various types of proteins, metabolites and other cell-signaling molecules. Designing an efficient genetic oscillator is one of the main focus of synthetic and systems biology. An efficient oscillator should be robust against parameter fluctuations and tunable across wide range of periods without compromising on the amplitude. We have developed a detailed theoretical framework of gene oscillators and using which we have derived the basic principles associated with robust and tunable oscillators. Independent gene oscillators can be coupled through either -OR- or -AND- type logics. Most of the coupled oscillators constructed so far seem to be -OR- type. Perturbation in one of the -OR-type oscillators does not change the overall period of the system (period-buffering) whereas the overall period moves towards the perturbed oscillator in case of -AND- type. Oscillators coupled through -OR- type logic are more sensitive to perturbations in the system parameters associated with the promoter state dynamics than -AND- type. Results presented here provide insights on the methodology of constructing robust and tunable synthetic gene oscillators.

## Introduction

Transcription factors (TFs) regulate the quantitative levels of several proteins inside a living cell [1]-[4]. TF networks present across various organisms ranging from prokaryotes to higher eukaryotes and consist of fundamental building blocks such as autoregulatory loops, cascades and single input modules, feed-forward and feedback loops, dense overlapping regulons and oscillatory loops [5]-[7]. Feedback loops act as bistable switches and feedforward loops have been shown to act as efficient filters for transient external signals [8], [10], [11]. Positive self-regulatory loops seem to play important roles in the maintenance of cellular memory [3] and subsequent reprogramming of the cellular states whereas negative auto regulatory loops have been shown [11] to speed up the response times against an external stimulus [8]-[10]. Oscillatory loops drive the developmental as well as mitotic cell-cycle dynamics [13] and circadian-rhythms [14], [15] associated with the intracellular concentration of various types of proteins, metabolites and other cell-signaling molecules. Understanding of the detailed dynamics of oscillatory loops associated with the TF networks is a central topic in biophysics, synthetic and systems biology.

The minimalist TF network model that can generate self-sustained oscillations is the well-known Goodwin-Griffith oscillator which has a single gene that codes for a TF protein that negatively auto-regulates its own transcription [16]-[18]. In this model the TF protein-product undergoes a one-step modification that yields the matured or active end-product and subsequently $n$ numbers of this end-product bind with the *cis*-regulatory modules (CRMs) of the associated promoter that in turn results in down-regulation. Here $n$ is the Hill coefficient associated with the cooperative type binding. Detailed studies on this minimalist model showed [17] that the inequality condition $n > 8$ is necessary to generate self-sustained oscillations in the levels of mRNA and protein. This result was obtained with the assumptions that the rate constants associated with the synthesis and decay of the protein and end-product are equal and the binding-unbinding of the end-product with the promoter is much faster than the rate of change in the synthesis and degradation of mRNA, protein and end-product. Further it was assumed that the decay of mRNA and protein product follows a first order type reaction.

It was realized later that the inequality condition $n > 8$ is unlikely [19] under *in vivo* conditions since the formation of such large multimeric protein complexes via pure three dimensional diffusion (3D) limited collisions (**Figure 1**) is almost an improbable event and several other modifications over the Goodwin-Griffith model were proposed to reduce the required value of $n$. Most of these modifications were mainly associated with the insertion of (a) a temporal delay in the negative auto-regulatory loop either explicitly as a time-delay in between the synthesis and binding of end-product at the promoter [19], [20] or implicitly via inserting additional reaction steps [19] in the formation of end-product that interacts with its own promoter and (b) a non-linear Michaelis-Menten type kinetics in the decay of mRNA and protein products despite of the first order type kinetics and (c) additional TF gene members in the negative feedback loop which again indirectly acts as temporal delay in the overall negative feedback. The delayed negative feedback may also be coherently or incoherently amplified [21]-[23] via the insertion of a positively regulated intermediate. Here the temporal delay is connected with the overall time that is required for the transport of fully transcribed mRNAs from nucleus to cytoplasm, post-translational modifications and subsequent transport of active TF proteins into the nucleus through 3D diffusion. When the decay of mRNA and protein product follows a Michaelis-Menten type kinetics then the Goodwin-Griffith (GG) oscillator seems to produce self-sustained oscillations [**19**] even for $n = 1$. The genetic oscillator module with three TF genes connected in a cyclic negative feedback loop is named as repressilator [**24**]. Though these motifs were shown to be oscillatory through deterministic and stochastic simulations, significant fraction of cells containing the constructs of these motifs were not showing any oscillations under *in vivo* experimental conditions. It was argued that it could be partially due to the noisy nature of intracellular environment [**18**], [**24**]. Here one should note that most of the simulation studies were performed with constant parameter values which may not be true under *in vivo* conditions. In this context it is essential to investigate how the oscillatory dynamics of these motifs reacts to perturbations in the system parameter values.

Most of the earlier studies on GG and other oscillator models assumed a quasi-equilibrium condition for the binding-unbinding dynamics of the negatively autoregulated TF proteins at their own promoters. This is mainly to reduce the four or higher dimensional Jacobian matrix associated with the non-linear system of differential rate equations into a three dimensional one to ease further analysis since there is an additional rate equation corresponding to the promoter state dynamics apart from the rate equations associated with mRNA, protein and end-product. However this assumption is valid [**8**], [**9**] only when the timescales associated with the synthesis and degradation of mRNAs and TF proteins are much slower than the timescales associated with the binding-unbinding of regulatory TFs at the respective promoters. Recent studies [**8**] on feedforward loops suggested that the binding-unbinding dynamics of TF protein at the promoter can be ignored only when the cellular volume $V_c$ (= volume of nucleus in case of eukaryotes) is comparable with that of the prokaryotes [**8**] such as *E. coli* ($V_c \sim 10^{-18}\,\mathrm{m}^3$) and the influence of the promoter state fluctuations on the overall dynamics of feedforward/feedback loops seems to significantly increase as the nuclear volumes increases as in eukaryotic cells across yeast to human. Further, the Michaelis-Menten type degradation kinetics associated with mRNA and protein is a valid assumption only when the concentrations of these species are much higher than the concentration of the corresponding nucleases and proteases. Nevertheless in most of the *in vivo* conditions, the intracellular levels of mRNA and protein of a particular TF gene will be much lesser than the corresponding levels of the non-specific nucleases and proteases. When the latter is true then the enzyme mediated decay of mRNA and protein will eventually follow a first order type kinetics. In this article, using a combination of theoretical and simulation tools (a) we develop a generalized theoretical framework of various types of genetic oscillators by explicitly incorporating the promoter state dynamics and other chemical reaction balances in detail. Using our detailed model (b) we identify and classify various critical control parameters and compute their physiological ranges which are required to generate self-sustained oscillations in the intracellular levels of mRNAs and transcription factor proteins and (c) explore various possibilities of coupling independent gene oscillators and fine-tuning the period of such coupled system. We further (d) demonstrate that by coupling two or more independent Goodwin-Griffith oscillators one can design oscillatory network architectures which are tunable and also robust against perturbations in system parameters.

## Results

### Theoretical framework of transcription factor gene oscillators

The Goodwin-Griffith oscillator consists of a negatively self-regulated gene (we denote it as TF gene A) which codes for a transcription factor protein (**Figure 2A**). We denote the cellular concentrations (mol/lit, M) of its mRNA as $m_a$, protein as $p_a$, the transformed end-product as $z_a$ and the complex of promoter with the end-product as $x_a$. Here the total cellular concentration of promoter is $d_{za}$ and the overall promoter state occupancy by the end-product will be $X_a = x_a/d_{za}$ where $X_a \in (0, 1)$. Though there is only one copy of the promoter by definition, we use a continuous type probability variable $X_a$ to describe promoter state occupancy mainly to account for its partially occupied status [**8**], [**9**]. The transcription and translation rates are denoted as $k_{ma}$ ($Ms^{-1}$) and $k_{pa}$ ($s^{-1}$) respectively. The first order decay rate constants corresponding to mRNA and TF protein are $\gamma_{ma}$ ($s^{-1}$) and $\gamma_{pa}$ ($s^{-1}$) respectively. The first order on- and off-rates associated with the transformation of protein into the matured end-product are denoted as $\lambda_{af}$ ($s^{-1}$) and $\lambda_{ar}$ ($s^{-1}$) and the corresponding dimensionless dissociation constant is $\lambda_a = \lambda_{ar}/\lambda_{af}$. The overall forward and reverse rate constants associated with the binding and unbinding of $n_a$ numbers of end-product molecules with the respective *cis*-regulatory modules (CRMs) of the promoter of TF gene A are $k_{af}$ ($M^{-n_a}s^{-1}$) and $k_{ar}$ ($s^{-1}$) and the corresponding dissociation constant is defined as $K_{arf} = k_{ar}/k_{af}$ ($M^{n_a}$). To simplify the analysis further we introduce the following scaling transformations to project the time and concentration variables into the dimensionless space.

$$\tau = \gamma_{pa}t; \; P_a = p_a/p_{as}; \; M_a = m_a/m_{as}; \; Z_a = z_a/p_{as}; \; X_a = x_a/d_{az} \tag{1}$$

In these equations $\tau$ denotes the dimensionless time that is measured as the number of lifetimes of the protein product of TF gene A and $P_a$, $M_a$, $Z_a$ and $X_a$ are respectively the dimensionless concentrations of protein, mRNA, end-product and promoter complexes. We also should note that $(P_a, M_a, Z_a$ and $X_a) \in (0, 1)$ by definition. Here the steady state values of mRNA and protein in the absence of negative self-regulation can be defined as follows [**8**], [**9**].

$$p_{as} = k_{ma}k_{pa}/\gamma_{ma}\gamma_{pa}; \; m_{as} = k_{ma}/\gamma_{ma} \tag{2}$$

We further transform the parameter associated with the multimerization of end-product and subsequent binding events as follows.

$$\mu_a = K_{arf}/p_{as}^{n_a}; \; K_{arf} = k_{ar}/k_{af} \tag{3}$$

Using the scaling transformations given by **Eqs (1-3)** one can write the deterministic rate equations corresponding to the temporal evolution of dimensionless concentration variables ($X_a$, $M_a$, $P_a$, and $Z_a$) over dimensionless time variable $\tau$ as follows.

$$v_a \, dX_a/d\tau = Z_a^{n_a}\left(1 - X_a\right) - \mu_a X_a$$
$$w_a \, dM_a/d\tau = \left(1 - X_a\right) - M_a$$
$$dP_a/d\tau = M_a - P_a - \sigma_a\left(P_a - \lambda_a Z_a\right) \tag{4}$$
$$\varepsilon_a \, dZ_a/d\tau = P_a - \left(\lambda_a + \kappa_a\right)Z_a - \chi_a\left(Z_a^{n_a}\left(1 - X_a\right) - \mu_a X_a\right)$$

The initial conditions are $\left(X_a, Z_a, M_a, P_a\right) = 0$ at $\tau = 0$. When ($v_a = 0$, $\sigma_a = 0$ and $\kappa_a = 0$), then this system reduces to the usual GG oscillator model for three concentration variables. Here we have defined the dimensionless ordinary perturbation parameter $\kappa_a = \gamma_{za}/\lambda_{af}$ where $\gamma_{za}$ $(\text{s}^{-1})$ is the first order decay rate constant associated with the protein end-product $Z_a$. Since $\gamma_{za} \gg \lambda_{af}$ will be true in most of the physiological conditions and $\kappa_a$ is an ordinary perturbation parameter one can assume $\kappa_a \approx 0$. When there is a dimerization of $z_a$ (we denote the dimer $z_a$-$z_a$ as $y_a$) as in case of *Lac* repressor system that has been constructed and studied in **Ref. 25** (negative-feedback-only model using *lacI* gene) then the first and last equations of **Eqs (4)** will be modified as follows.

$$v_a \, dX_a/d\tau = Y_a^{n_a}\left(1 - X_a\right) - \mu_a X_a$$
$$\varepsilon_a \, dZ_a/d\tau = P_a - \left(\lambda_a + \kappa_a\right)Z_a - \sigma_{ya}\left(Z_a Z_a - \lambda_{ya} Y_a\right) \tag{5}$$
$$\varepsilon_{ya} \, dY_a/d\tau = Z_a Z_a - \lambda_{ya} Y_a - \chi_{ya}\left(Y_a^{n_a}\left(1 - X_a\right) - \mu_a X_a\right)$$

Here various parameters associated with the dimerization of end-product of TF protein A and subsequent assembly of this end-product at the own promoter are defined as follows.

$$\chi_{ya} = p_{as}^{n_a-1} d_{az} k_{af}/\lambda_{fya} \; ; \; \varepsilon_{ya} = \gamma_{pa}/\lambda_{fya} p_{as} \; ; \; \lambda_{ya} = \lambda_{rya}/p_{as}\lambda_{fya} \; ; \; \sigma_{ya} = p_{as}\lambda_{fya}/\lambda_{fa}$$

Here we have defined $Y_a = y_a/p_{as}$ and $\lambda_{fya}$ $(\text{M}^{-1} \text{ s}^{-1})$ is the forward rate constant associated with the dimerization reaction and $\lambda_{rya}$ $(\text{s}^{-1})$ is the corresponding reverse rate constant. One should note that for a fully functional Lac repressor system the Hill coefficient will be $n_a = 4$ since an octamer of Lac repressor protein (which is a dimer) is involved in the overall looping of DNA that results in strong repression of *lacI*. The system of **Eqs (4)** is completely characterized by the following set of dimensionless parameters.

$$v_a = \gamma_{pa}/p_{as}^{n_a} k_{af} \; ; \; w_a = \gamma_{pa}/\gamma_{ma} \; ; \; \varepsilon_a = \gamma_{pa}/\lambda_{af} \; ; \; \sigma_a = \lambda_{af}/\gamma_{pa} \; ; \; \chi_a = p_{as}^{n_a-1} d_{az} k_{af}/\lambda_{af}$$

Here one should note that the parameters $\left(\mu_a, \chi_a, v_a\right)$ are functions of $n_a$ that can be simplified by assuming an *in vivo* protein level as $p_{as} = 1$. To simplify the analysis further we can classify these dimensionless control parameters into Group I, II and III. Group I consists of $\left(v_a, w_a, \varepsilon_a\right)$ which are all singular type perturbation parameters since they multiply the first order derivative terms. One should note that this set of parameters directly controls the dynamics of changes in the cellular concentrations of active promoter, mRNA and end-product respectively. Group II consists of $(\sigma_a, \chi_a, \kappa_a)$ those are ordinary type perturbation parameters. In Group II, $\sigma_a$ controls the coupled dynamics associated with the concentrations of TF protein A and its end-product whereas $\chi_a$ controls the coupled dynamics of changes in the concentrations of end-product and it's binding with the promoter sequence. The lifetime of end-product is controlled by $\kappa_a$. One should note that almost all the earlier studies assumed that $(\sigma_a, \chi_a, \kappa_a) = 0$. Group III consists of the equilibrium and promoter affinity parameters $(\lambda_a, \mu_a)$. In this $\lambda_a$ controls the equilibrium associated with the formation of end-product and $\mu_a$ controls the equilibrium associated with the binding of $n_a$ molecules of end-product with CRMs of the promoter of TF gene A.

## Biophysical modeling of promoter state dynamics

The total time required to initiate transcription consists of at least two different components viz. the time required (proportional to $1/k_{af}$) for the assembly of $n_a$ numbers of TFs at the respective CRMs of promoter and the time required for subsequent looping of DNA and subsequent distal action of TFs on RNAPII-promoter complex. The time component associated with the looping and distal action along with the time required for elongation and termination steps are included in the definition of transcription rate (the total time required for transcribing a full length mRNA will be equal to $1/k_{ma}$).The kinetics of interaction of $n_a$ molecules of end-product with the sequentially located CRMs can occur in two different possible ways namely binding of the full-fledged complex of $n_a$ molecules of end-product (pathway II) or sequential assembly of the monomers of end-product at the corresponding *cis*-regulatory DNA-binding sites similar to that of a combinatorial binding of TFs with CRMs (pathway I) as in eukaryotic systems (**Figure 1**). Though the pathway I resembles a $(n_a+1)^{th}$ order reaction it is still an operable one since the length scale of the genomic DNA is much higher than the combinatorial binding TF proteins. Binding of $n_a$ numbers of transcription factors in a sequential manner or $n_a$-mer of end-product leads to looping of the DNA segment that is present in between promoter and CRMs of TF gene A that results in the spatial or distal communication between the end-product present at CRMs and the already formed RNAPII-promoter complex which in turn activates (positive) or deactivates (negative) the initiation of transcription depending on the type of self-regulation [1], [3], [4], [26], [27]. In case of activation or positive regulation, the combinatorial transcription factors bound at CRMs enhance the initiation of transcription by strengthening the RNAPII-promoter interactions through their distal action (positive arrows in **Figure 2**) whereas in case of negative regulation, the RNAPII-promoter complex will be destabilized by the combinatorial TFs present at CRMs (negative arrows in **Figure 2**). Here the destabilization of RNAPII-promoter complex may be through the formation of stem and loop structures. In prokaryotes, these types of up and down regulations generally do not involve recruitment or combinatorial binding of several TFs and the regulator transcription factor directly influences the RNAP-promoter interactions as in case of negatively self-regulated oscillatory motifs constructed with a *lac*-repressor gene. Here binding of *lac*-repressor at the Operator sequence directly destabilizes RNAP-promoter complex that in turn lead to the down regulation of transcription [1], [3]. The total time $\tau_{d,n_a}$ required for the formation of a full-fledged $n_a$-mer via 3D diffusion-controlled collisions and subsequent binding with the *cis*-regulatory sites can be calculated as follows.

$$\tau_{d,n_a} = n_a\tau_{s,1} + \sum_{i=1}^{n_a}\tau_{d,i} = n_a\tau_{s,1} + \tau_{d,1}\sum_{i=1}^{n_a}i/(1+i) = n_a\tau_{s,1} + \tau_{d,1}\left(n_a + 1 - \Psi\left(n_a+2\right) - \gamma_e\right) \qquad (6)$$

In this equation $n_a\tau_{s,1}$ is the time required for the searching and binding of the entire $n_a$-mer at the corresponding CRMs on DNA (for a monomer it will be $\tau_{s,1}$) via a combination of 1D and 3D diffusion, $\tau_{d,1}$ is the time that is required for the formation of a dimer of the end-product through 3D diffusion under *in vivo* conditions, $\Psi\left(n_a\right) = d\ln\Gamma\left(n_a\right)/dn_a$ where $\Gamma\left(n_a\right)$ is Gamma function and $\gamma_e = 0.5772157..$ is the Euler-Mascheroni constant. Here $\tau_{d,1} = 10^{-3}/16\pi RD_T N_A C_Z$ (s) is the minimum possible 3D diffusion controlled bimolecular collision time inside the cellular volume where $R$ is average radius of the monomers of end-product, $C_Z$ (mol/lit) is the concentration of end-product inside the cellular volume, $N_A$ is the Avogadro's number, $D_T = k_B T/6\pi\varphi R$ (m²s⁻¹) is the 3D diffusion coefficient associated with the dynamics of monomers of end-product in aqueous medium where $k_B$ is the Boltzmann constant, $T$ is the absolute temperature $(K)$, $\varphi$ is the viscosity of the medium. In the calculation of $\tau_{d,1}$, we have assumed that the reaction radius between two monomers is $\sim 2R$. Since the overall maximum radius of the $m$-mer will be $\sim mR$, we find that $\tau_{d,m} \propto m/(1+m)$ and subsequently the total time that is required to form a $n_a$-mer in a sequential manner via 3D diffusion will be given by the sum $\tau_{d,1} \sum_{m=1}^{n_a} m/(1+m)$. We should note that this is the maximum possible search time since we have assumed a maximum possible radius for the $n_a$-mer complex and also we have not considered the possibility of formation of the final $n_a$-mer through non-sequential and random pathways and the steric factor associated with the multimerization reaction. The total search-time that is required by the monomer of end-product to find its cognate site on DNA is defined as $\tau_{s,1} = N(\tau_L + \tau_{ns})/L$ where the overall 1D sliding time is defined as $\tau_L = L^2/6x_d$ [28]. In this calculation we have assumed that the end-product searches for its binding sites on DNA via a combination of 1D and 3D diffusion controlled collision routes. Monomers of the end-product undergo at least $N/L$ numbers of cycles of 3D diffusion mediated association that is followed by 1D scanning and dissociation where $N$ is the size of the genomic DNA (base-pairs, bps) and $L$ (bps) is the average 1D sliding-length between non-specific 3D association and dissociation. Here $\tau_L$ is the time that is required by the monomers of end-product to scan $L$ bps of the genomic DNA via 1D diffusion along the DNA chain, $x_d$ (bps$^2$s$^{-1}$) is the 1D diffusion coefficient associated with the sliding of monomers on DNA (this will be scaled down to $x_d/n_a$ for $n_a$-mer) and $\tau_{ns}$ is the time that is required for non-specific binding of end-products with the genomic DNA via 3D collisions under in vivo conditions. When all the $n_a$ monomers of end-product search for their binding sites on DNA in a parallel manner, then the total time [26] that is required ($\tau_{s,n_a}$) for all these $n_a$ monomers to assemble at the sequentially located *cis*-acting elements can be derived from the theory of combinatorial binding of transcription factors [27]-[29] with DNA as follows.

$$\tau_{s,n_a} = N(\tau_L n_a^\alpha + \tau_{ns})/L \approx \tau_{s,1} n_a^\alpha \qquad (7)$$

From this equation we find that the 1D scanning time increases with the number of monomers $n_a$ in a power law manner as $n_a^\alpha$ where typical value of the exponent seems to be $\alpha \sim 2/5$ [26]. From **Eqs (6)** and **(7)** one can compute the following ratio.

$$\theta_{n_a} = (\tau_{d,n_a}/\tau_{s,n_a}) = n_a^{1-\alpha} + (\tau_{d,1}/\tau_{s,1})(n_a + 1 - \Psi(n_a+2) - \gamma_e)n_a^{-\alpha} \qquad (8)$$

From the theory of site-specific DNA-protein interactions we find that $(\tau_{d,1}/\tau_{s,1}) \approx 10^2$ for $n_a = 1$ [26]-[29] which suggests that $\theta_{n_a} \geq 10^2$ for all values of $n_a$. **Eqs (7)** and **(8)** suggest the pathway I is more efficient than pathway II. This means that though the diffusion limited multimerization of the monomers of the end-product is not a reasonable assumption for large values of $n_a$, direct assembly of the monomers of the end-products of TFs on the sequentially located *cis*-regulatory DNA binding sites via a combination of 1D and 3D diffusion controlled routes can be still a reasonable assumption even for higher values of $n_a$. In this context we can replace the combinatorial binding of $n_a$ numbers of TFs with CRMs of template DNA with a single step $(n_a+1)^{th}$ order reaction as given by the first equation in **Eqs (4)**. One should note that unlike the prokaryotic systems, most of the eukaryotic promoters are activated through a combinatorial binding of several TFs at the corresponding CRMs [27]. This observation suggests that GG oscillator can be still a feasible model that can be used to generate limit-cycle oscillations in the cellular levels of negatively self-regulated TF proteins especially in eukaryotic systems.

## Steady-state analysis of Goodwin oscillator

The system of **Eqs (4)** has a fixed point $P_a = \eta_a$ which is a real solution of the following polynomial equation of the order $(n_a+1)$.

$$\mu_a \Big/ \Big(\mu_a + \big(\eta_a/(\lambda_a + \kappa_a)\big)^{n_a}\Big) - \eta_a \beta_a = 0; \ \beta_a = \big(1 + \sigma_a \kappa_a/(\lambda_a + \kappa_a)\big) \tag{9}$$

The steady state values of other concentration variables ($Z_a$, $M_a$ and $X_a$) can be calculated using the fixed point $\eta_a$ as follows.

$$Z_{as} = \eta_a/(\kappa_a + \lambda_a); \ X_{as} = (1 - \eta_a \beta_a); \ M_{as} = \eta_a \beta_a$$

Using the Jacobian matrix evaluated around the equilibrium point $P_a = \eta_a$, the linearized form of the system of **Eqs (4)** near this equilibrium point can be written as follows.

$$\frac{d}{d\tau}\begin{pmatrix} X_a \\ M_a \\ P_a \\ Z_a \end{pmatrix} \approx \begin{pmatrix} -g & 0 & 0 & A_a/v_a \\ -1/w_a & -1/w_a & 0 & 0 \\ 0 & 1 & -(1+\sigma_a) & \sigma_a \lambda_a \\ \delta & 0 & 1/\varepsilon_a & -c \end{pmatrix}\begin{pmatrix} X_a \\ M_a \\ P_a \\ Z_a \end{pmatrix} \tag{10}$$

Here we have defined various matrix elements as follows.

$$g = \mu_a/\beta_a \eta_a v_a; \ c = (\lambda_a + \kappa_a + \chi_a A_a)/\varepsilon_a; \ \delta = \chi_a \mu_a/\beta_a \eta_a \varepsilon_a; \ A_a = n_a \mu_a (\lambda_a + \kappa_a)(1 - \beta_a \eta_a)/\eta_a$$

The coefficients associated with the characteristic polynomial $Y^4 + rY^3 + sY^2 + tY + u = 0$ (**PI**) of the Jacobian matrix defined in **Eqs (10)** can be written as follows.

$$r = g + c + (1 + \sigma_a) + 1/w_a$$

$$s = g(1 + \sigma_a) + cg + c(1 + \sigma_a) + \big(g + c + (1 + \sigma_a)\big)\big/w_a - \sigma_a \lambda_a/\varepsilon_a - A_a \delta/v_a$$

$$t = \big((1 + \sigma_a)g + cg + c(1 + \sigma_a)\big)\big/w_a + cg(1 + \sigma_a) - \big(A_a \delta(1 + \sigma_a) + 1/w_a\big)/v_a - \sigma_a \lambda_a \big(g + 1/w_a\big)/\varepsilon_a$$

$$u = A_a/v_a \varepsilon_a w_a + \big(cg(1 + \sigma_a) - A_a \delta(1 + \sigma_a)/v_a\big)\big/w_a - \sigma_a \lambda_a g/w_a \varepsilon_a$$

From the Routh-Hurwitz criterion for a biquadratic polynomial [30] we find that the equilibrium point of the system $P_a = \eta_a$ will be stable only when all the following inequality conditions are true. In other words there may be limit-cycle oscillations around the steady state only when any of these inequality conditions is not true.

$$r > 0; \ (rs - t) > 0; \ (rs - t)t - r^2 u > 0; \ u > 0 \tag{11}$$

Here the first inequality condition $r > 0$ will be always true since $\eta_a \propto 1/\beta_a$ and subsequently we find $A_a > 0$. The second inequality condition $(rs - t) > 0$ can be reversed only when either $s < 0$ or $rs < t$ for $s > 0$. When the third inequality in **Eqs (11)** is not true then the biquadratic polynomial can have a complex root such that $y_{Re} \pm iy_{Im}$ with positive real-part (here $y_{Re} \geq 0$ $y_{Im} \geq 0$) that results in the generation of limit-cycle oscillations of the concentration of TF protein A and the period of such oscillations [29]-[32] will be given by $\tau_p = 2\pi/y_{Im}$. This means that the period of GG oscillator can be modified by tuning the lifetime ($\gamma_{pa}$) of the protein product of TF gene A (since $\tau = \gamma_{pa}t$) though the value of $y_{Im}$ is a function of other Group I parameters ($w_a, v_a, \varepsilon_a$) which are in turn linearly depend on $\gamma_{pa}$. Since the Hill coefficient term $n_a$ presents only in the coefficient terms $s, t$ and $u$, the third inequality in **Eqs (11)** can be reversed by increasing the value of $n_a$ for any set of Group I, II and III parameters. This means that the parameter space that is required to generate oscillations can be expanded by increasing the value of $n_a$. Inequality conditions given by **Eqs (11)** for a stable motion of the dynamical system of **Eqs (4)** can be directly derived from the following Routh table ($RT_{GG}$) [30] corresponding to the biquadratic polynomial (**PI**).

$$R_{GG} = \begin{bmatrix} 1 & s & u & Y^4 \\ r & t & 0 & Y^3 \\ s - t/r & u & 0 & Y^3 \\ t - ru/(s - t/r) & 0 & 0 & Y \end{bmatrix} \tag{12}$$

When $y_{Im} = 0$ then the steady state solution will be either asymptotically stable or unstable depending on the values of real-parts. From **Eqs (10-11)** we find that the system will be inconsistent near the fixed point both at very large as well as small values of Group I type parameters as $(v_a, w_a, \varepsilon_a) \to 0$ or $\infty$. This means that there exists a critical range of these parameters to generate limit-cycle oscillations in the cellular levels of TF protein A. One should note that Group I parameters appear in the denominator of definitions of various coefficients of the characteristic polynomial (**PI**) which means that the period of oscillations will increase proportionately with respect to an increase in these parameters since we find $y_{Im} \propto 1/(v_a, w_a, \varepsilon_a)$. Contrasting from Group I, there exist critical or threshold values of Group II type ordinary perturbation parameters $(\chi_a, \sigma_a, \kappa_a)$ below which the oscillations occur. As in case of Group I type parameters, there exist a critical range of values of $(\lambda_a, \mu_a)$ Group III to generate limit-cycle oscillations. The critical value of Hill coefficient $n_a = {}_C n_a$ for a given set of parameters can be iteratively calculated by numerically solving the third inequality in **Eqs (11)** at various values of $n_a$. When the dynamics of promoter state occupancy is much faster than the rate of change in the concentrations of other variables then one can set $v_a \approx 0$ and the system of **Eqs (4)** reduces to the following form.

$$w_a \, dM_a / d\tau = \mu_a \big/ \big(\mu_a + Z_a^{n_a}\big) - M_a$$
$$dP_a / d\tau = M_a - P_a - \sigma_a \big(P_a - \lambda_a Z_a\big) \tag{13}$$
$$\varepsilon_a \, dZ_a / d\tau = P_a - \big(\lambda_a + \kappa_a\big) Z_a$$

Most of the earlier studies on GG oscillator consider **Eqs (13)** as the base model however with the conditions such that $\big(\sigma_a, \kappa_a\big) = 0$. Using detailed numerical simulations we will show later that this assumption is reasonably invalid. The corresponding Jacobian matrix around the steady state $P_a = \eta_a$ can be written as follows.

$$\frac{d}{d\tau}\begin{pmatrix} M_a \\ P_a \\ Z_a \end{pmatrix} \approx \begin{pmatrix} -1/w_a & 0 & -A'_a/w_a \\ 1 & -(1+\sigma_a) & \sigma_a \lambda_a \\ 0 & 1/\varepsilon_a & -(\lambda_a+\kappa_a)/\varepsilon_a \end{pmatrix}\begin{pmatrix} M_a \\ P_a \\ Z_a \end{pmatrix} \tag{14}$$

Here we have defined $A'_a = n_a \beta_a \big(1 - \beta_a \eta_a\big)\big(\lambda_a + \kappa_a\big)$ and the characteristic polynomial associated with this equation is $Y^3 + r'Y^2 + s'Y + t' = 0$ (**PII**) where the coefficients are defined as follows.

$$r' = 1 + \sigma_a + \big(\lambda_a + \kappa_a\big)/\varepsilon_a + 1/w_a$$
$$s' = \big(\lambda_a + \kappa_a\big)\big(w_a + 1\big)/w_a \varepsilon_a + \big(1 + \sigma_a\big)/w_a + \kappa_a \sigma_a / \varepsilon_a \tag{15}$$
$$t' = A_a / w_a \varepsilon_a + \big(\lambda_a + \kappa_a\big(1 + \sigma_a\big)\big)/w_a \varepsilon_a$$

The Routh-Hurwitz [30] condition required by the system of **Eqs (13-14)** to generate limit-cycle oscillations will be $\big(r's' - t'\big) < 0$. Upon solving this inequality for the Hill coefficient $n_a$, the expression for the critical value of $n_a$ that is required to generate limit cycle oscillations can be obtained as follows.

$$_C n_a = \big(w_a \varepsilon_a r's' - \big(\lambda_a + \kappa_a\big(1 + \sigma_a\big)\big)\big)\big/ \beta_a \big(1 - \beta_a \eta_a\big)\big(\lambda_a + \kappa_a\big) \tag{16}$$

Here one should note that the term $\eta_a$ in the right hand side of this equation is still a function of $n_a$ and $\mu_a$ and the following limiting conditions exist.

$$\lim_{\mu_a \to 0} \eta_a = \big(\mu_a \big(\lambda_a + \kappa_a\big)^{n_a} \big/ \beta_a\big)^{1/(n_a+1)}; \ \lim_{\mu_a \to \infty} \eta_a = 1/\beta_a; \ \lim_{n_a \to \infty} \eta_a = 1/\beta_a \tag{17}$$

**Eqs (15-17)** suggest that strong binding of the end-product ($\mu_a \Rightarrow 0$) at the promoter of TF gene A is required along with the conditions such as $(\kappa_a, \sigma_a) = 0$, and $(\lambda_a, w_a, \varepsilon_a) = 1$ to decrease the required critical Hill coefficient towards the minimum possible value as $_c n_a \approx 9$. When there is an additional dimerization step as in **Eqs (4)** and **(5)** then the resulting characteristic polynomial of the Jacobian matrix will be of fifth order as $Y^5 + rY^4 + sY^3 + tY^2 + uY + m = 0$ (**PIII**) and the Routh criterion that is required to generate oscillations can be written as follows.

$$K\left(tL - rK\right) - mL^2 < 0; \ K = u - m/r; \ L = s - t/r$$

## One-to-one dual feedback oscillators

One can extend these scaling ideas for one-to-one negative feedback oscillator or toggle switches (**Figure 2B1**) and repressilator models (**Figure 2C1**). In case of one-to-one negative feedback oscillator $n_a$ number of end-product molecules of TF gene A bind with the *cis*-regulatory elements associated with the promoter of TF gene B and subsequently down-regulates whereas $n_b$ number of end-product molecules of TF gene B down-regulate the promoter of TF protein A upon binding with the corresponding *cis*-regulatory elements (**Figure 2B1**). The set of differential rate equations associated with the two TFs one-to-one feedback system can be written in the dimensionless form as follows.

$$v_h \, dX_h/d\tau = Z_q^{n_q}\left(1 - X_h\right) - \mu_h X_h$$
$$w_h \, dM_h/d\tau = \left(1 - X_h\right) - M_h$$
$$\rho_h \, dP_h/d\tau = M_h - P_h - \sigma_h\left(P_h - \lambda_h Z_h\right)$$
$$\varepsilon_h \, dZ_h/d\tau = P_h - \left(\lambda_h + \kappa_h\right)Z_h - \chi_h\left(Z_h^{n_h}\left(1 - X_q\right) - \mu_q X_q\right)$$

(18)

In these equations the subscripts will be such that when $h = a, b$ then $q = b, a$ where $(a, b)$ denote the TF genes A and B respectively. One should note that the Hill coefficients associated with the binding of the end-product of TF A at the promoter of TF B and end-product of TF gene B at the promoter of TF gene A are $n_a$ and $n_b$ respectively and in general $n_a \neq n_b$. Here we have defined various other dimensionless variables and parameters as follows.

$$\tau = \gamma_{pa}t; \ P_h = p_h/p_{hs}; \ M_h = m_h/m_{hs}; \ Z_h = z_h/p_{hs}; \ X_h = x_h/d_{hc}; \ h = a, b$$
$$v_h = \gamma_{pa}/p_{qs}^{n_q}k_{hf}; \ w_h = \gamma_{pa}/\gamma_{mh}; \ \varepsilon_h = \gamma_{pa}/\lambda_{hf}; \ \rho_h = \gamma_{pa}/\gamma_{ph}; \ \sigma_h = \lambda_{hf}/\gamma_{ph}; \ \chi_h = p_{hs}^{n_h-1}d_{qz}k_{qf}/\lambda_{qf}$$
$$\lambda_h = \lambda_{hr}/\lambda_{hf}; \ \mu_h = K_{hrf}/p_{hs}^{n_h}; \ K_{hrf} = k_{hr}/k_{hf}; \ \beta_h = 1 + \sigma_h\kappa_h/\left(\lambda_h + \kappa_h\right); \ \kappa_h = \gamma_{zh}/\lambda_{hf}$$

In these definitions for $h = a, b$ one needs to set $q = b, a$. The steady state solutions $\left(\eta_b, \eta_a\right)$ to the coupled set of **Eqs (18)** corresponding to this two TFs system with respect to the scaled protein levels can be given as follows.

$$P_{as} = \eta_a = e^y; \ P_{bs} = \eta_b = \lambda_a^{n_a/n_b}\left(\mu_a\left(1 - \beta_a\eta_a\right)/\eta_a\beta_a\right)^{1/n_b}$$

(19)

The steady state values of other dynamical variables can be calculated using the steady state values of the protein products $\eta_a$ and $\eta_b$ as follows.

$$Z_{hs} = \eta_h \big/ (\kappa_h + \lambda_h); \; X_{hs} = (1 - \eta_h \beta_h); \; M_{hs} = \eta_h \beta_h; \; h = a,b \qquad (20)$$

Here $y$ in **Eqs (19)** is the real root of $y(n_a n_b - 1) - n_b \ln \Phi + \ln \left( \mu_a \lambda_a^{n_a} \beta_a^{-1} \left(1 - \beta_a e^y\right) \right) = 0$ where we have defined the function $\Phi$ as,

$$\Phi = \mu_b \lambda_b^{n_b} \beta_b^{-1} \left(1 - \beta_b \lambda_a^{n_a/n_b} \left( \mu_a \left(1 - \beta_a e^y\right) \big/ \beta_a \right)^{1/n_b} e^{-y/n_b} \right).$$

**Eqs (18)** have three possible steady state solutions viz. $(\eta_b = \eta_a)$, $(\eta_b < \eta_a)$ and $(\eta_b > \eta_a)$. Under identical values of all the parameters such as ($\mu_a = \mu_b$, $n_a = n_b$ and so on) we find the unstable steady state solution of the two TFs system as $\eta_b = \eta_a$ where $0 \leq (\eta_b = \eta_a) \leq 1$. This means that the limit-cycle oscillations around this unstable steady state can occur only when the values of all the control parameters and initial conditions are identical with respect to both the TF genes A and B. Using the eighth order characteristic polynomial of the Jacobian matrix associated with the linearized form of **Eqs (18)** (Methods section) near the steady state values $(\eta_b = \eta_a)$, one can numerically derive the conditions for the occurrence of oscillations from the Routh-Hurwitz criterion. When the values of the control parameters are such that ($\mu_a \neq \mu_b$ or $n_a \neq n_b$ and so on) or there is a transient perturbation in the values of these parameters or initial conditions, then the oscillating system will be unstable and driven to any one of the stable steady state solutions as either $\eta_b < \eta_a$ or $\eta_b > \eta_a$ through asymptotic spirals. For example when $n_b \gg n_a$ or $\mu_b \gg \mu_a$ then the stable steady-state solution will be ($\eta_a \approx 0$, $\eta_b \approx 1$). These results suggest that contrasting from GG oscillator model the identical two-TF feedback system cannot generate self-sustained oscillations in the presence of stochastic noise. One can also construct one-to-one feedback oscillator via coupling two independent GG oscillators. Here these independent TF oscillators A and B can be coupled via either A-OR-B (**Figure 2B2**) or A-AND-B (**Figure 2B3**) type logics. One can consider various types of regulatory combinations associated with these network architectures. The combinations in A-OR-B type coupling can be denoted as 'AsAc-BsBc' where 'As' and 'Bs' are the types of self-regulation of TF genes A and B respectively whereas 'Ac' and 'Bc' are the types of their cross-regulation on each other. Each type of regulation can be either 'P' or 'N' where 'P' denotes positive type and 'N' denotes the negative type regulation. Using these notations one can denote the configuration given in **Figure 2B2** as NN-NN type, **Figure 2B4** as NN-PP type and **Figure 2B5** as NP-NP type. The configurations given in **Figures 2B4** and **2B5** are the well-studied robust dual-feedback oscillators [**25**], [**33**]-[**35**]. Similarly one can consider various possibilities in A-AND-B type architectures. Noting the symmetry of regulation we find three possible types as P-P, N-N and N-P out of which only N-N will be a robust oscillator. The configuration given in **Figure 2B3** is an N-N type. When the coupling is via A-OR-B type logic then both the promoters of TF genes A and B will be independently down-regulated upon binding of protein end-products of both the TF genes A and B at the respective *cis*-regulatory elements associated with each promoter and the first and last equations in **Eqs (18)** will be modified as follows.

$$v_{hq}\, dX_{hq}/d\tau = Z_q^{\,n_{hq}}\left(1-X_h\right) - \mu_{hq}X_{hq}; \; X_h = \sum_{m=a,b} X_{hm}; \; h,q=a,b$$

$$\varepsilon_h\, dZ_h/d\tau = P_h - \left(\lambda_h+\kappa_h\right)Z_h - \chi_{hq}\left(Z_h^{\,n_{hq}}\left(1-X_q\right)-\mu_{hq}X_{hq}\right) - \chi_{hh}\left(Z_h^{\,n_{hh}}\left(1-X_h\right)-\mu_{hh}X_{hh}\right)$$

(21)

In these equations for each value of subscript '$h$' the subscript '$q$' will take $a$, $b$ and there are totally four equations associated with the overall promoter state dynamics. Various modified parameters in **Eqs (21)** are defined as follows.

$$\mu_{hq} = K_{hqrf}/p_{hs}^{\,n_{hq}}; \; K_{hqrf} = k_{hqr}/k_{hqf}; \; v_{hq} = \gamma_{pa}/p_{hs}^{\,n_{hq}}k_{hqf}; \; \chi_{hq} = p_{qs}^{\,n_{hq}-1}d_{qz}k_{hqf}/\lambda_{hf}$$

The steady state solutions corresponding to the modified equations can be obtained by numerically solving the following set of coupled polynomial equations.

$$\mu_{hh}\mu_{hq}\Big/\left(\mu_{hh}\mu_{hq}+\mu_{hh}Z_{qs}^{\,n_{hq}}+\mu_{hq}Z_{hs}^{\,n_{hh}}\right)-\beta_h\eta_h=0; \; h=a,b; \; q=b,c$$

(22)

Here $Z_{hs}$ is defined as in **Eqs (20)**. When $\mu_{hq}=\bar{\mu}$ and $\beta_h=1$ then we can calculate the steady state protein levels from the set of polynomial **Eqs (22)** as $\eta_h=\left(\lambda_h+\kappa_h\right)e^y$ where $y$ is the real root of $y\left(n_a+1\right)-\ln\left(\bar{\mu}\left(1-e^y\right)-e^{y\left(n_a+1\right)}\right)=0$. When the GG oscillators A and B are coupled through A-AND-B type logic then the dimer ($y_d$) of both end-products ($z_a$-$z_b$) will be the key regulating molecule that binds at the *cis*-acting elements associated with the promoters of both TF genes A and B however with different Hill coefficients ($n_h$) and subsequently down-regulate them. The respective modified differential rate equations corresponding to the dimerization and binding of dimer at the promoters of TF genes A and B can be written as follows.

$$\varepsilon_d\, dY_d/d\tau = Z_a Z_b - \phi_d Y_d - \sum_{h=a,b}\chi_h\left(Y_d^{\,n_h}\left(1-X_h\right)-\mu_h X_h\right)$$

$$v_h\, dX_h/d\tau = Y_d^{\,n_h}\left(1-X_h\right)-\mu_h X_h; \; h=a,b$$

$$\varepsilon_h\, dZ_h/d\tau = P_h - \left(\lambda_h+\kappa_h\right)Z_h - \chi_{dh}\left(Z_a Z_b - \phi_d Y_d\right)$$

(23)

The modified and new parameters and variables in **Eqs (23)** are defined as follows.

$$Y_d = y_d/p_{as}; \; \varepsilon_d = \gamma_{pa}/p_{bs}\lambda_{df}; \; \chi_{dh} = \lambda_{df}p_{qs}/\lambda_{hf}; \; \phi_d = \lambda_{dr}/\lambda_{df}p_{bs}; \; \chi_h = p_{hs}^{\,n_h-1}d_{hz}k_{hf}/\lambda_{hf}p_{qs}$$

In the definition of $\chi_h$ for $h=a$, $b$ one needs to substitute $q=b$, $a$. Here $\lambda_{df}$ (M⁻¹s⁻¹) and $\lambda_{dr}$ (s⁻¹) are the forward and reverse rate constants associated with the diffusion limited dimerization reaction between the protein end-products of TF gene A and B. The corresponding steady state solutions to **Eqs (24)** can be written as follows.

$$Y_{ds} = Z_{hs}Z_{qs}/\lambda_{hq}; \; Z_{hs} = P_{hs}/\left(\lambda_h+\kappa_h\right); \; X_{hs} = Y_{ds}^{\,n_h}\Big/\left(\mu_h+Y_{ds}^{\,n_h}\right); \; P_{hs}=\eta_h$$

(24)

Here $\eta_h$ is the solution to the set of following polynomial equations.

$$\mu_h \Big/ \left( \mu_h + \left( \left( \eta_h \eta_q \big/ (\lambda_h + \kappa_h)(\lambda_q + \kappa_q) \right) \big/ \phi_d \right)^{n_h} \right) - \beta_h \eta_h = 0; \; h = a, b; \; q = b, a \tag{25}$$

In this set of equations we need to set $q = b$ for $h = a$ and for $h = b$ we need to set $q = a$. The parameters associated with dual feedback oscillators are summarized in **Table 2**.

## Three gene repressilator type oscillators

Similar to **Eqs (18)** one can write the set of differential rate equations associated with the three TFs repressilator model as follows (**Figure 2C1**).

$$\begin{aligned}
v_h \, dX_h / d\tau &= Z_s^{n_s} (1 - X_h) - \mu_h X_h \\
w_h \, dM_h / d\tau &= (1 - X_h) - M_h \\
\rho_h dP_h / d\tau &= M_h - P_h - \sigma_h (P_h - \lambda_h Z_h) \\
\varepsilon_h \, dZ_h / d\tau &= P_h - (\lambda_h + \kappa_h) Z_h - \chi_h \left( Z_h^{n_h} (1 - X_q) - \mu_q X_q \right)
\end{aligned} \tag{26}$$

Here the subscripts will be such that $(h = a, b, c; \; s = c, a, b; \; q = b, c, a)$ where $(a, b, c)$ denotes respectively TF gene A, B and C in a cyclic $(h, s, q)$ manner and the variables as well as various control parameters are defined as in case of **Eqs (18)** and generally $n_a \neq n_b \neq n_c$. Similar to **Eqs (19)** one can derive the steady state solutions with respect to the scaled protein levels for three TFs system as follows.

$$\begin{aligned}
P_{as} &= \eta_a = e^y \\
P_{bs} &= \eta_b = \left( \eta_a \mu_c \lambda_c^{n_c} \beta_a \left( 1 - \beta_c \left( \mu_a \lambda_a^{n_a} (1 - \beta_a \eta_a) / \eta_a \beta_a \right)^{1/n_c} \right) \Big/ \mu_a \lambda_a^{n_a} (1 - \beta_a \eta_a) \beta_c \right)^{1/n_c n_b} \\
P_{cs} &= \eta_c = \left( \mu_a \lambda_a^{n_a} (1 - \beta_a \eta_a) / \eta_a \beta_a \right)^{1/n_c}
\end{aligned} \tag{27}$$

In this equations $y$ is the real root of $y(n_a n_b n_c + 1) - n_c n_b \ln \Delta + n_c \ln \phi - \ln \psi = 0$ where we have defined various other terms as,
$\Delta = \mu_b \lambda_b^{n_b} \beta_b^{-1} \left( 1 - \beta_b e^{(\phi n_c - \psi + y)/n_c n_b} \right); \; \psi = \mu_a \lambda_a^{n_a} \beta_a^{-1} \left( 1 - \beta_a e^y \right); \; \phi = \mu_c \lambda_c^{n_c} \beta_c^{-1} \left( 1 - \beta_c e^{-y/n_c} \psi^{1/n_c} \right).$

Using these steady state values of protein products $\eta_h$, one can write the steady state solutions to other dynamical variables can be written similar to **Eqs (20)** as follows.

$$Z_{hs} = \eta_h / (\kappa_h + \lambda_h); \; X_{hs} = (1 - \eta_h \beta_h); \; M_{hs} = \eta_h \beta_h; \; h = a, b, c \tag{28}$$

Under identical values of all the control parameters such as ($\mu_a = \mu_b = \mu_c$, $n_a = n_b = n_c$ and so on), the system reaches the steady state as $(\eta_a = \eta_b = \eta_c)$ which is an unstable fixed point since even a small perturbation in the parameter values or initial conditions will drive the system towards a stable limit-cycle. As depicted in **Figures 2C2** and **2C3** one can also construct the repressilator type model by cyclically coupling three independent GG oscillators A/B/C through -AND- or -OR- type logical gates as we have constructed in **Figures. 2B2-3**. When the type of interaction is through -AND- type logic, then the $z_a$-$z_b$ dimer down-regulates TF gene B, $z_b$-$z_c$ dimer down-regulates TF gene C and the $z_c$-$z_a$ dimer down-regulates TF gene A. The set of modified differential equations corresponding to the configuration that is given in **Figure 2C3** can be written as follows.

$$\varepsilon_{hk}\, dY_{hk}/d\tau = Z_h Z_k - \lambda_{hk} Y_{hk} - \varphi_{hk}\left(Y_{hk}^{n_k}\left(1 - X_k\right) - \mu_k X_k\right);\ Y_{hk} = y_{hk}/p_{hs}\,;\ h = a,b,c;\ k = b,c,a$$

$$\nu_h\, dX_h/d\tau = Y_{kh}^{n_h}\left(1 - X_h\right) - \mu_h X_h\,;\ h = a,b,c;\ k = c,a,b \qquad (29)$$

$$\varepsilon_h\, dZ_h/d\tau = P_h - \left(\lambda_h + \kappa_h\right)Z_h - \chi_{hk}\left(Z_h Z_k - \lambda_{hk} Y_{hk}\right) - \chi_{hq}\left(Z_h Z_q - \lambda_{hq} Y_{hq}\right);\ k = b,c,a;\ q = c,a,b$$

The steady state solutions can be obtained by numerical methods from the following set of algebraic equations.

$$Y_{hks} = Z_h Z_k/\lambda_{hk}\,;\ X_{hs} = Y_{khs}^{n_k}/\left(\mu_h + Y_{khs}^{n_h}\right);\ \mu_h/\left(\mu_h + \left(Z_{hs} Z_{ks}/\lambda_{hk}\right)^{n_h}\right) - \beta_h \eta_h = 0 \qquad (30)$$

In these equations for $h = a,\ b,\ c$ one needs to set $k = c,\ a,\ b$ and various new and modified parameters are defined as follows.

$$\lambda_{hk} = \lambda_{hkr}/\lambda_{hkf}\, p_{ks}\,;\ \varphi_{hk} = k_{hkf}\, d_{zk}\, p_h^{n_k - 1}/\lambda_{hkf}\, p_{ks}\,;\ \varepsilon_{hk} = \gamma_{pa}/p_{ks}\, k_{hkf}\,;\ \chi_{hk/q} = \lambda_{hk/qf}\, p_{k/qs}/\lambda_{hf}$$

When the type of interaction is through -OR- type logic as depicted in **Figure 2C2** then the end products of both TF genes A and B can independently down-regulate TF gene B and the end products of TF genes B and C can independently down-regulate TF gene C and so on. The set of modified differential equations associated with such system will be similar to that of **Eqs (21-23)** where the indices will be extended for three TF genes A/B/C. One can also consider a fully interconnected network of TF genes A/B/C. In these configurations the self-regulated promoters of each TF gene A/B/C will be negatively regulated by the end-products of the remaining two other TF genes. Here the mode of overall combinatorial interactions among these regulating end-products and the corresponding promoters can be either through -AND- or -OR- type logics as represented by the dashed lines in **Figures 2C2-3**. In case of -OR- type logical gate, the negatively self-regulated promoter of TF gene A will also have *cis*-regulatory binding sites for the end-products of both TF genes B and C and so on. One can write the modified set of differential equations associated with such fully interconnected configuration (**Figure 2C2**) as follows.

$$\nu_{hq}\, dX_{hq}/d\tau = Z_q^{n_{hq}}\left(1 - X_h\right) - \mu_{hq} X_{hq}\,;\ X_h = \sum\nolimits_{m = a,b,c} X_{hm}\,;\ h,q,k = a,b,c$$

$$\varepsilon_h\, dZ_h/d\tau = \left\{ \begin{array}{l} P_h - \left(\lambda_h + \kappa_h\right)Z_h - \chi_{hq}\left(Z_h^{n_{hq}}\left(1 - X_q\right) - \mu_{hq} X_{hq}\right) - \chi_{hh}\left(Z_h^{n_{hh}}\left(1 - X_h\right) - \mu_{hh} X_{hh}\right) \\ -\chi_{hk}\left(Z_h^{n_{hk}}\left(1 - X_k\right) - \mu_{hk} X_{hk}\right) \end{array} \right\} \qquad (31)$$

In the first one of **Eqs (31)**, there will be three equations for each promoter and there are totally nine equations associated with the overall promoter state dynamics. In the second set of three equations as well as in the associated parameters for each value the subscript '*h*' from the set (*a*, *b*, *c*), the subscripts *q* and *k* will take the remaining values. This means that when *h* = *a* then *q* = *b* and *k* = *c* and so on. Various modified parameters in **Eqs (31)** are defined as follows.

$$\mu_{hq} = K_{hqrf} \big/ p_{hs}^{n_{hq}} \; ; \; K_{hqrf} = k_{hqr} \big/ k_{hqf} \; ; \; v_{hq} = \gamma_{ph} \big/ p_{hs}^{n_{hq}} k_{hqf}$$

$$\chi_{hh} = p_{hs}^{n_{hh}-1} d_{hz} k_{hhf} \big/ \lambda_{hf} \; ; \; \chi_{hq} = p_{qs}^{n_{hq}-1} d_{qz} k_{hqf} \big/ \lambda_{hf} \; ; \; \chi_{hk} = p_{ks}^{n_{hk}-1} d_{kz} k_{hkf} \big/ \lambda_{hf}$$

The steady state solution to **Eqs (31)** needs to be obtained by numerically solving the following set of equations.

$$\mu_{hh} \mu_{hq} \mu_{hk} \Big/ \Big( \mu_{hh} \mu_{hq} \mu_{hk} + \mu_{hh} \mu_{hq} Z_{ks}^{n_{hk}} + \mu_{hh} \mu_{hk} Z_{qs}^{n_{hq}} + \mu_{hq} \mu_{hk} Z_{hs}^{n_{hh}} \Big) - \beta_h \eta_h = 0; \; Z_{hs} = \eta_h \Big/ \big( \lambda_h + \kappa_h \big)$$

In case of fully interconnected configuration through -AND- type logic that is depicted in **Figure 2C3** (with dashed lines), the complex $z_a$-$z_b$-$z_c$ will be the key regulating molecule that binds with the promoters of all the three TF genes A/B/C and down-regulate them. Similar to **Eqs (23)** one can write the modified set of differential equations corresponding to repressilator configuration that is fully interconnected through -AND- type logic as follows.

$$\varepsilon_d \, dY_d / d\tau = Z_a Z_b Z_c - \phi_d Y_d - \sum_{k=a,b,c} \chi_k \Big( Y_d^{n_k} \big( 1 - X_k \big) - \mu_k X_k \Big)$$

$$v_h \, dX_h / d\tau = Y_d^{n_h} \big( 1 - X_h \big) - \mu_h X_h; \; h = a,b,c \qquad (32)$$

$$\varepsilon_h \, dZ_h / d\tau = P_h - \big( \lambda_h + \kappa_h \big) Z_h - \chi_{dh} \big( Z_a Z_b Z_c - \lambda_d Y_d \big)$$

Here we have defined $Y_d = y_d / p_{as}$. The steady state solutions to this equation can be obtained by solving the following set of polynomial equations.

$$\mu_h \Big/ \Big( \mu_h + \big( Z_{hs} Z_{qs} Z_{ks} / \lambda_d \big)^{n_h} \Big) - \beta_h \eta_h = 0; \; Z_{ms} = \eta_m \Big/ \big( \lambda_m + \kappa_m \big) \qquad (33)$$

Here various modified and new parameters are defined as follows.

$$\varepsilon_d = \gamma_{pa} \big/ p_{cs} p_{bs} \lambda_{df} \; ; \; \chi_{dh} = \lambda_{df} p_{qs} p_{ks} \big/ \lambda_{hf} \; ; \; \phi_d = \lambda_{dr} \big/ \lambda_{df} p_{cs} p_{bs} \; ; \; \chi_h = p_{as}^{n_k-1} d_{hz} k_{hf} \big/ \lambda_{df} p_{bs} p_{cs}$$

In these equations similar to **Eqs (31)** for each value the subscript '*h*' from the set (*a*, *b*, *c*), the subscripts *q* and *k* will take the remaining values. This means that when *h* = *a* then *q* = *b* and *k* = *c* and so on for other values.

## Perturbation-responses of various oscillators

Sample trajectories and phase portraits of GG oscillator for $v_a = \big( 0, 2 \times 10^{-4} \big)$ are shown in **Figures 3A1-3** and **4A1-3**. Irrespective of the type of initial conditions and magnitude of the control parameters, the trajectories always start with an overshoot of protein production that is followed by asymptotic spirals towards a stable limit cycle. This seems to be an inherent property of negatively self-regulated loops [9]. **Figures 3B1-4** and **4B1-4** suggest that there exists an optimum range of Group I parameters $v_a = (4-12) \times 10^{-4}$ and $(w_a, \varepsilon_a) \in (0.2, 1.8)$ at which the critical Hill coefficient ($_C n_a$) that is required to generate self-sustained oscillations is a minimum which is in turn strongly dependent on the promoter state occupancy parameter $\mu_a$. This optimum range is also dependent on the values of other Group II and III parameters. The optimum range of the conversion parameter seems to be $\lambda_a \in (0.6-1.4)$. Results suggest that strong binding conditions $\mu_a \sim 2 \times 10^{-4}$ ($v_a \neq 0$) and $\mu_a < 10^{-12}$ ($v_a = 0$) are required to minimize the value of critical Hill coefficient with respect to changes in Group I type parameters. The minimum achievable values of critical Hill coefficients seems to be $_C n_a = 6$ ($v_a \neq 0$) and $_C n_a = 9$ ($v_a = 0$). When there is an additional dimerization step as described in **Eqs (4-5)** corresponding to the negative-feedback-only (NFO) model considered in **Ref. 25**, the minimum achievable critical Hill coefficient seems to be $_C n_a = 3$. One should note that in the *Lac I* oscillatory system the effective Hill coefficient is $n_a = 4$ since four dimers of *lac I* end-products involved in the overall negative feedback. Numerical analysis of this NFO model system using the physiological range of parameters as given in **Table 1** suggests that the period of oscillator can be well tuned by changing the promoter state affinity $\mu_a$ of the repressor without compromising the amplitude much as shown in **Figure 3C1**.

**Figure 4B4** shows the strong influence of $\sigma_a$ on the critical $_C n_a$ which means that the approximation ($\sigma_a = 0$) as in case of most of the earlier studies on various genetic oscillators is not a valid one. At the critical Hill coefficient, the period as well as amplitude of oscillations are strongly dependent on the Group I parameters as shown in **Figures 5A1-3**. These results also demonstrate how the oscillator responds to temporal perturbations is Group I parameters. As we have shown in the theory section, the period of oscillations increases with increase in the Group I parameters whereas the amplitude seems to decrease as the value of Group I parameters increase. One should note that square of period of oscillation is inversely proportional to the total energy of an oscillator whereas the total energy is directly proportional to the square of amplitude. This means that the total energy of a GG oscillator can be fine-tuned by perturbing the Group I parameters. The Goodwin-Griffith oscillator seems to abruptly enter into the modified limit-cycle orbit upon introducing the perturbation and relax back much faster upon removal of perturbation in the parameter $v_a = 0$ rather than perturbations in other parameters $(w_a, \varepsilon_a)$. In the latter cases, as shown in **Figures 5A2-3** the relaxation of oscillator to the original orbit upon removal of perturbation seems to be through slow asymptotic spirals. **Figure 5B1** suggest that the period of oscillations increases monotonically with respect to increase in the value of Group I parameters as we have predicted in the theory section. When $v_a \neq 0$ then there exists a range of $w_a \in (0.3, 1)$ at which the period of limit cycle oscillations and the required $_C n_a$ are almost independent of changes in $w_a$. **Figure 5B2** shows that when $v_a \neq 0$ then the period of oscillations linearly increases as $\sigma_a$ increases whereas it linearly decreases with increase in $\sigma_a$ when $v_a = 0$.

The dual feedback motif (**Figures 2B1-3**) is an adaptable one that can act as a toggle switch as well as an oscillator depending on the type of configuration. As we have shown in the theory section, the configuration depicted in **Figure 2B1** requires identical values of all the control parameters as well as initial conditions to generate coupled as well as synchronized oscillations. Particularly this configuration can efficiently act as a toggle switch since the fixed point $\eta_a = \eta_b$ is an unstable one and even small perturbations in the parameters or initial conditions is enough for the system to exit from the synchronized limit cycle oscillations around this unstable fixed point and subsequently move towards any one of the two stable steady states. The configurations given in **Figures 2B2-3** can act as coupled oscillators. Sample trajectories and phase portraits of one-to-one coupled oscillators corresponding to the configuration given in **Figure 2B1** are shown in **Figures 6A1-6.** The minimum achievable value of the critical Hill coefficient that is required to generate self-sustained oscillations around the unstable fixed point ($\eta_a = \eta_b$) of dual feedback oscillator seems to be $_cn_h = 5$ that is closer ($_cn_h = 6$) to the critical value corresponding to GG oscillator. Variations of critical Hill coefficient with respect to changes in combination of different groups of control parameters are shown in **Figures 6B1-8.** Results suggest that the minimum value of critical Hill coefficient that is required to generate self-sustained oscillations can be achieved only when $\mu_h \leq 2 \times 10^{-4}$ and $(w_h, \varepsilon_h) \in (0.5, 2)$. **Figure 6B1** also suggests that the inequality condition $v_h > 4 \times 10^{-6}$ is required for oscillations. From **Figure 6B2** we find that $_cn_h$ is also independent on the changes in the ordinary perturbation parameter $\chi_h$. However it is strongly dependent on $\sigma_h$ and the condition $\sigma_h \leq 0.3$ is required to achieve the minimum value of critical $_cn_h$. Results suggested that when there are identical perturbations in the given control parameter, then the one-to-one coupled oscillator (**Figure 2B1**) behaves similar to that of GG oscillator. That is to say the period of oscillations increases and amplitude decreases with an increase in Group I parameters. Here the identical perturbations are such that for the parameter $w_h$ we have $w_h \Rightarrow w_h + \delta_{wh}$ where $w_a = w_b$ prior to perturbation and the magnitude of perturbation is such that $\delta_{w_a} = \delta_{w_b}$. When any of these two conditions fails, then the system will be driven towards the corresponding stable steady state.

Upon receiving a transient pulse of perturbation or imbalance in the control parameters the dual feedback oscillator exits from the limit-cycle orbit with a time-delay ($\tau_{del}$) and subsequently reaches one of the stable steady-states via asymptotic spirals. Here the target steady state is dependent on the type of disproportion in the parameter values among TF genes A and B. For example when the perturbation is from $\mu_a = \mu_b$ towards $\mu_a < \mu_b$ then the target steady-state will be $\eta_a > \eta_b$ since the binding of TF end-product B at the promoter of TF gene A is stronger than the binding of end-product A at the promoter of TF gene B. It seems that the value of this time-delay is dependent on the extent of disproportion ($\pi_k$, where the subscript 'k' denotes the control parameter under consideration such as $\mu_h$) as well as duration of the perturbation ($\tau_w$) in control parameters or initial conditions associated with the TF genes A and B. Here the percentage of disproportion or imbalance with respect to the parameter $k_b$ that is associated with TF gene B is defined as $\pi_k = 100|k_a - k_b|/k_b$. **Figure 6C** shows the variation of the time-delay with respect to changes in the extent of disproportion ($\pi_\mu$) in the control parameter $\mu_h$ and duration of perturbation $\tau_w$. It seems that $\tau_{del}$ approaches zero independently upon increase in both $\tau_w$ and $\pi_k$. Further simulation results suggest that the time-delay $\tau_{del}$ is independent on the time ($\tau_{pulse}$) at which the perturbation in the control parameter is introduced into the system.

Contrasting from the configuration given in **Figure 2B1**, the limit-cycle orbits of the coupled oscillators depicted in **Figures 2B2-3** are robust against transient imbalances in the control parameters. The minimum achievable value of the critical Hill coefficient seems to be $_c n_h = 4$ for the oscillator with A-OR-B type logic (**Figure 2B2**) whereas $_c n_h = 2$ for the coupled oscillators with A-AND-B type logic (**Figure 2B3**). Results suggest that the limit cycle orbit of coupled oscillators with A-AND-B and A-OR-B type logics are stable one. When there are temporal perturbations in Group I parameters associated with one of the Goodwin oscillators (TF gene A/B) then the other unperturbed oscillator responds to the changes in the behavior of the perturbed oscillator depending on the type of logical coupling between them. As shown in **Figures 7A1-2**, **B1-2** and **C1-2** in case of A-OR-B coupling an increase in the magnitude of Group I parameters associated with one of the oscillators A/B does not change the period of the entire system of oscillators (period-buffering) though there is a decrease in the amplitude of the oscillator that is perturbed in $(w_h, \varepsilon_h)$. The decrease in the amplitude might be partially owing to the period-buffering effect. In case of A-AND-B type logical coupling, increase in the magnitude of Group I parameters $(w_h, \varepsilon_h)$ increases the period of oscillations and decreases the amplitude of the entire system of oscillators that includes both TF genes A/B. **Figures 8A1-4** suggest that an increase in the parameter $v_h$ of one of the oscillators initially increases the amplitude of other oscillator to a maximum which then decreases later. Perturbations in Group I parameters $(w_h, v_h)$ associated with one of the oscillators A/B also results in a phase-shift in cases of both A-AND-B and A-OR-B type logical couplings as shown in **Figures 7-8A1-2** and **B1-2**. Whereas perturbation in $\varepsilon_h$ affects only the amplitude and does not affect the phases of the coupled oscillators A and B as shown in **Figures 7-8C1-2**. Here one should note that in case of A-OR-B type coupling the parameter $v_h$ will be split into $v_h \Rightarrow (v_{hh}, v_{hk})$ where we have the indices $(h, k) = a, b$ as given in **Eqs (21)**. Above results corresponding to A-OR-B type logical coupling with respect to changes in the parameter $v_h$ are valid only when the temporal perturbations are the same for a given promoter of TF gene A/B. This means that for TF gene A (here we have subscript $k = a$) the extent of perturbation should be the same for both $v_{aa}$ and $v_{ab}$ while the set of parameters associated with the TF gene B ($k = b$) remains unperturbed. Here one should note that $v_{hh}$ controls the dynamics associated with the binding of end-product of TF gene 'H' at its own promoter whereas $v_{hk}$ controls the dynamics associated with the binding of end-product of TF gene 'K' at the promoter of TF gene 'H' as we have shown in **Eqs (21)**. When there are perturbations in only one of these two split parameters (as $v_a \Rightarrow (v_{aa}, v_{ab})$) then the coupled system of oscillators seems to be dynamically unstable and also produces modulated beats as shown in **Figures 7A3-4**. The period of such beats increases as the imbalance in the set of split parameters $v_{hk}$ increases as shown in **Figures 7A5-6**. These dynamical instabilities as well as beats abruptly disappear once the perturbations in $v_{hk}$ are removed. Whereas the system of coupled oscillators relaxes back to the initial unperturbed limit-cycle orbit through asymptotic spirals upon removal of perturbations in case of Group I control parameters $\left(w_h, \varepsilon_h\right)$. Results from **Figures 7** and **8** suggest that coupled oscillators with -AND- type logic are more robust against promoter state perturbations than the -OR- type coupling. Period of a network of oscillators can be easily fine-tuned by manipulating merely one of the oscillators when the mode of coupling is via -AND- type.

Sample trajectories and phase portraits of a repressilator configuration given by **Figure 2C1** are shown in **Figure 9A1-5**. The minimum achievable value of the critical Hill coefficient for the repressilator seems to be $_cn_h = 2$ similar to that of a one-to-one feedback oscillator with A-AND-B type logical coupling. As we have shown in the theory section, the steady state fixed point is more stable when the parameters and initial conditions are identical for all the TF genes A/B/C. When there is a transient perturbation in the control parameters then the system leaves the steady state and enters into a stable limit-cycle orbit through asymptotic spirals with a time delay $\tau_{del}$ as shown in **Figures 9A4**. Here the magnitude of this time delay seems to be directly proportional to the extent of imbalance or disproportions in the parameter values as shown in **Figures 9A4**. Perturbation in the control parameter $v_h$ associated with any one of the TF genes A/B/C results in the decrease of amplitude of the perturbed as well as the one that regulates it. However perturbation in $v_h$ does not affect the period of oscillations of the entire system of TF genes as shown in **Figures 9B1-2**. This means that when $v_a$ is increased then the amplitudes of oscillations of TF genes A and C decrease whereas the amplitude of B is not affected. Perturbation in the control parameters $\left(w_h, \varepsilon_h\right)$ associated with any one of the TF genes A/B/C decreases the amplitude of oscillations of the TF gene that is regulated by the end-product of the perturbed TF gene and increases the amplitude as well as width of oscillations of the TF gene that is regulating the perturbed gene. This means that when $\left(w_a, \varepsilon_a\right)$ increases then the amplitude of oscillations of TF genes A and B decreases whereas the width and amplitude of TF gene C increases. Further results show that an increase in $\left(w_h, \varepsilon_h\right)$ of any one of the oscillators increases the period of oscillations of the entire system of oscillators as shown in **Figures 9B3-4**.

Sample trajectories and phase plane portraits associated with the configuration given in **Figure 2C2** are shown in **Figures 10A1-3** and **B1-4**. Contrasting from three TF genes repressilator model (**Figure 2C1**) the configurations given in **Figures 2C2-3** do not require any asymmetry in the values of control parameters or initial conditions to trigger the stable oscillations. When the mode of coupling of TF genes A/B/C of GG oscillators is through -OR- type logic then in the presence of identical values of all the sets of control parameters the TF genes A/B/C oscillate in a synchronized manner with respect to period and amplitude. When there is a perturbation in set of the control parameters $v_{hq}$ (here $v_h$ will be split into $\left(v_{hh}, v_{hq}\right)$ for each promoter) associated with any one of the TF genes A/B/C then there are at least three different phases of responses. In the first phase, as shown in **Figures 10B1-2** the system tries to resist the perturbation by keeping the synchronized limit-cycle orbit intact whereas in the second phase the system becomes unstable and chaotic whose magnitude depends on the extent of perturbation. Upon removal of perturbation, in the third phase the system enters into new asynchronous limit-cycle orbit with stable phase differences among the TF gene oscillators. When there is a perturbation in one of the split parameters $\left(v_{hh}, v_{hq}\right)$, then as shown in **Figure 10B1** the second phase will have several repeating elements of resistance and instability. Perturbation in the control parameters $\left(w_h, \varepsilon_h\right)$ seems to have similar effects which are evident from **Figures 10B3-4**. Contrasting from these results, the oscillator depicted in **Figure 2C3** seems to be more robust against changes in the Group I control parameters and also they return back to the initial coherent type limit-cycle orbit upon removal of perturbations as shown in **Figures 10C1-2**. Sample trajectories of fully interconnected configurations given in **Figures 2C2-3** (with dashed lines) are shown in **Figures 11A1-3** and **B1-3**. These results suggests that the fully interconnected three TF genetic oscillator will be more stable against perturbations in the critical control parameters when the mode of coupling is through -AND- type logic than the -OR- type logic.

Tuning capabilities of A-AND-B (**Figure 2B3**) and A-OR-B (**Figure 2B2**) type coupled dual feedback oscillators are demonstrated in **Figures 12A-D** and **13A-D**. These results show that A-AND-B type coupled oscillators can be tuned by changing the promoter state binding parameter $\mu_h$ efficiently without conceding on the amplitude of oscillations. Increase in $\mu_h$ monotonically decreases the amplitude (and increases the period) of A-OR-B type coupled oscillators. Whereas A-AND-B type oscillator shows two distinct regions of responses with respect to changes in $\mu_h$ namely a responsive region and nonresponsive region. For the settings in **Figure 12B**, the A-AND-B system responds to the perturbations in $\mu_h$ when $\mu_h < 10^{-5}$. When $\mu_h > 10^{-5}$ then changes in $\mu_h$ will not affect the period of oscillations much. On the other hand the amplitude (measured in terms of $P_h/P_{hs}$) of A-AND-B oscillator is $\geq 1$ in a wide range of $\mu_h$ as well as $w_h$ values. Further A-OR-B type coupled oscillators behaves similar to that of single GG oscillator with respect to changes in the perturbation of Group I parameter $w_h$. In case of both types of coupled oscillators the amplitude seems to be inversely proportional to the critical $_c n_h$ rather than the period of oscillations as it is apparent from **Figures 12C** and **D** and there exists an optimum value of Group I parameter $w_h$ at which the amplitude of oscillations is a maximum. **Figure 13A-D** shows that increase in Group I parameter $v_h$ increases the period of oscillations and decreases the critical $_c n_h$ in both -AND- and -OR- type coupled oscillators. However the period of -AND-type oscillator seems to be more robust against changes in $v_h$ than -OR- type coupled dual feedback oscillator. Contrasting from these there exist a cutoff value of $\varepsilon_h \sim 2$ (for the simulation settings in **Figure 13**) beyond which the amplitude of oscillations is practically zero in both types of coupled oscillators that is apparent from **Figures 13C** and **D**.

## Discussion

Transcription factor gene oscillators play critical roles in driving cell-cycle to circadian rhythms. Here we have identified the critical control parameters associated with self-sustained oscillations of such oscillators and classified them into Groups I, II and III depending on their functionality. Group I parameters control the intracellular dynamics of synthesis and degradation of various molecules associated with regulated TF gene (**Figure 1**). The parameter $w_h$ of Group I describes the strength of coupling between the rate of degradation of mRNAs and the rate of degradation of corresponding TF proteins. We should note that transcription and translation of various TF genes of prokaryotes are simultaneously taking place well within the cytoplasm whereas in case of eukaryotes the transcription is taking place inside the nucleus and the synthesized pre-mRNA transcripts need to be spliced within the nucleus and then transported to cytoplasm after other post-transcriptional modifications through nuclear pores for translation. These differences in the cellular architecture demands higher lifetimes ($1/\gamma_{mh}$) for eukaryotic mRNAs than the prokaryotic ones which results in the general observation that the values of $w_h$ associated with various genes in prokaryotes are lower than eukaryotes genes [36], [37]. It seems that in case of yeast, the genome-wide values of $w_h$ varies from 0.1 to 1 with a median of ~0.3 [36]. These observations suggest that the naturally occurring or synthetic oscillatory motifs in the transcription networks should operate well within this range of $w_h \in (0.1,1)$ with different median values depending on the type of organism. For most of the bacterial genes we find that $w_h \sim 0.1$ [8], [9]. The parameter $v_a$ describes the dynamics of binding-unbinding of the end-product molecule with the promoter of TF gene A. Large values of $v_a$ indicate slower changes in the residence state of promoters whereas small values indicate the faster dynamics of the promoter state towards the equilibrium. Earlier studies suggested [8], [9] a typical physiological value of $v_a$ as $v_a \sim 10^{-4}$ for a prokaryotic self-regulatory systems ($n_a = 1$) such as the one in *E. coli*. Whereas in case of eukaryotic systems such as yeast and human its value will be scaled up respectively to $v_a \sim 10^{-3}$ and $v_a \sim 10^{-2}$ owing to dilution in the local concentrations of various molecules upon increase in the nuclear volumes. One should note that yeast nucleus is ~$10^1$ times higher than *E. coli* cell whereas human cell nucleus is ~$10^2$ times higher than *E. coli* cell. The parameter $\varepsilon_a$ describes the dynamics associated with of conversion of the protein product to end-product towards the equilibrium. This conversion reaction can be either a first-order conformational transition or pseudo first-order chemical modification of the protein product via an additional catalyst. Here $\varepsilon_a$ also acts as an indirect delay parameter that in turn relates the protein decay rate with the conversion rate. One should note that the binding-parameter $\mu_a$ is the central one that connects the entire Group I type singular perturbation parameters. By definition $\mu_a$ is inversely proportional to the affinity of the end-product towards the promoter sequence. Along with $\mu_a$ the ordinary parameters $\left( \sigma_a, \lambda_a, \kappa_a \right)$ decide the steady state value of TF protein. Further one should note that $\mu_a$ (or $\mu_h$ in general) is the only parameter that can be externally modified in a working model and all the others are fixed system parameters which can be modified only at the time of designing the oscillatory module. In case of *Lac* based oscillators $\mu_a$ can be modified through changing the concentration of IPTG [25] which is a gratuitous inducer of *lacI* gene.

Understanding the design principles associated with the genetic oscillator is one of the central topics in synthetic and systems biology. An efficient synthetic oscillator should be robust against transient perturbations in the control parameters and fluctuations in the promoter state occupancies. The period of oscillators should be easily tunable without compromising the amplitude. Stricker *et.al* in **Ref. 25** suggested that the robustness of the dual-feedback oscillators depicted in **Figures 2B4-5** can be further increased by the explicit delay owing to the oligomerization steps associated with the end-products of TF genes A/B (here gene A can be *lacI* and gene B can be *araC*). Further studies by Tsai *et.al* in **Ref. 35** suggested that unlike pure negative feedback systems, inclusion of a positive feedback loop within the negative feedback oscillators (**Figure 2B4-5**) can increase the robustness and tunable nature of the original oscillator without compromising on the overall amplitude. Here the positively auto regulated TF gene acts as a booster for the overall protein level of the negatively self-regulated oscillatory TF protein. In this context one can also consider a positive coupling of a positively auto-regulated TF protein K (booster) with N-N or NN-NN type dual feedback oscillators as depicted in **Figure 2B6**. Upon analyzing various classes of oscillators depicted in **Figure 2** one can conclude that the effects of perturbations in the control parameters and promoter state fluctuations will be minimum when the mode of coupling among network of oscillators is through -AND- logic. From **Figures 12** and **13** we find that the A-AND-B coupled dual feedback oscillator can be externally tuned by modifying the promoter state binding $\mu_h$ without conceding on the amplitude of oscillations. The inherent chaotic response of the -OR- type coupled oscillators could be one of the reasons why the oscillations of most of the synthetic oscillators disappeared after certain period of time [18], [24] under *in vivo* conditions. Nevertheless the overall period of network of oscillators coupled through -OR- type logic are more resistant (period-buffering) to perturbations in the control parameters than the oscillators coupled through -AND- type logic. One should note that the temporal perturbations in the system parameters ultimately result in perturbations in the intra cellular concentrations of protein end-products of coupled TF genes that in turn are looped back into the system. In this context the -OR- type coupled genes respond to perturbations in an additive way whereas those genes coupled through -AND- type logic respond in a multiplicative way. This could be the reason why the period of oscillations of -OR- type coupled oscillators are more resistant to perturbations than -AND- type coupled oscillators. This means that the period of network of oscillators coupled through -AND- type logic can be more easily tuned by perturbing any one of the coupled oscillators. Depending on the circuit functionality one can also chose a combination of both the types of coupling. So far we have assumed a copy number of TF genes as one ($d_{hz} = 1$). This may not be true in the bacterial based synthetic circuits constructed on plasmids since the plasmid copy number will be more than one in most of the times which can lead to further complexity. For example, a plasmid copy number of two for a NFO model that was constructed in **Ref. 25** can mimic a combination of oscillators depicted in **Figure 2B2** and **2B3** with TF gene A = B since it can mimic both -AND- as well as -OR- type coupled GG oscillators. This could be one of the possible reasons for observing stable and robust oscillations with such constructs [25] over several bacterial generations.

In summary, in this paper we have considered various types of transcription factor oscillators by explicitly incorporating the promoter state dynamics and other chemical reaction balances in detail. Using our detailed model we have identified and classified various critical control parameters and numerically obtained their physiological and optimum ranges to generate self-sustained oscillations in the intracellular levels of mRNAs and transcription factor proteins. We further derived the basic design principles associated with robust and tunable gene oscillators. We have further demonstrated that by coupling two or more independent Goodwin-Griffth oscillators via -OR- or -AND- type logics one can construct genetic-oscillators which are fine-tunable and also robust against perturbations in the system parameters. When there is a perturbation in one of the -OR- type coupled oscillators, then the overall period of the system remains constant whereas in case of -AND- type of coupling the overall period of the system moves towards the perturbed oscillator. Though there is a period-buffering, the oscillators coupled through -OR- type logic seems to be more sensitive to perturbations in the parameters associated with the promoter state dynamics than -AND- type.

## Materials and Methods

We use Euler type numerical scheme [**8**], [**9**], [**38**] to integrate the set of differential rate equations corresponding to various types of oscillatory loops. For example in case of **Eqs (4)** which describe the Goodwin-Griffith model, the numerical integration scheme can be written as follows.

$$X_{a,i+1} = X_{a,i} + \left(Z_{a,i}^{n_a}\left(1-X_{a,i}\right) - \mu_a X_{a,i}\right)\Delta\tau/v_a$$

$$M_{a,i+1} = M_{a,i} + \left(\left(1-X_{a,i}\right) - M_{a,i}\right)\Delta\tau/w_a$$

$$P_{a,i+1} = P_{a,i} + \left(M_{a,i} - P_{a,i} - \sigma_a\left(P_{a,i} - \lambda_a Z_{a,i}\right)\right)\Delta\tau \qquad (34)$$

$$Z_{a,i+1} = Z_{a,i} + \left(P_{a,i} - \left(\lambda_a + \kappa_a\right)Z_{a,i} - \chi_a\left(Z_{a,i}^{n_a}\left(1-X_{a,i}\right) - \mu_a X_{a,i}\right)\right)\Delta\tau/\varepsilon_a$$

Initial and boundary conditions are $\left(X_a, M_a, P_a, Z_a\right) = 0$ and $\left(X_a, M_a, P_a, Z_a\right) \leq 1$ respectively. The scaled time step $\Delta\tau$ should be chosen such that it captures the dynamics of all the variables including the dynamics of promoter state occupancies (first one in **Eqs (34)**). We divide the total scaled simulation time $T$ into $N$ equal intervals such that $\Delta\tau = T/N$. For simulation purpose we set $\Delta\tau = 10^{-5}$ and the corresponding $\Delta t = 0.03$s for a lifetime $1/\gamma_{pa} \sim 60$ mins. We use Newton's method [**38**] to find the fixed point solutions to steady state equations. Here to obtain the solution to a nonlinear algebraic equation $f\left(x\right) = 0$ one uses the iterative scheme $x_{i+1} = x_i - f\left(x_i\right)/f'\left(x_i\right)$. For example, the iterative numerical scheme to obtain the fixed point solution $P_a = \eta_a$ to **Eqs (9)** particularly for $n_a > 4$ can be written as follows.

$$\eta_{a,i+1} = \eta_{a,i} + \left(\mu_a - \beta_a\eta_{a,i}\left(\mu_a + \psi_a\eta_{a,i}^{n_a}\right)\right)\Big/\left(\mu_a + \left(n_a+1\right)\eta_{a,i}^{n_a}\psi_a\right); \ \psi_a = \left(\lambda_a + \kappa_a\right)^{-n_a} \qquad (35)$$

In this numerical scheme we set the initial guess value of the fixed point solution as $\eta_{a,i=0} = 0$ and the tolerance limit for stopping iteration as $\left|\eta_{a,i} - \eta_{a,i+1}\right| \leq 10^{-5}$. We can use the following computational workflow. For example, in case of one-to-one dual feedback oscillator one needs to construct the eighth dimensional Jacobian matrix associated with **Eqs (18)** as follows.

$$J_{OTO} = \begin{pmatrix}
-g_a/v_a & 0 & 0 & 0 & 0 & 0 & 0 & A_a/v_a \\
-1/w_a & -1/w_a & 0 & 0 & 0 & 0 & 0 & 0 \\
0 & 1/\rho_a & -\left(1+\sigma_a\right)/\rho_a & \sigma_a\lambda_a & 0 & 0 & 0 & 0 \\
0 & 0 & 1/\varepsilon_a & -c_a/\varepsilon_a & \chi_a g_b/\varepsilon_a & 0 & 0 & 0 \\
0 & 0 & 0 & A_b/v_b & -g_b/v_b & 0 & 0 & 0 \\
0 & 0 & 0 & 0 & -1/w_b & -1/w_b & 0 & 0 \\
0 & 0 & 0 & 0 & 0 & -1/\rho_b & -\left(1+\sigma_b\right)/\rho_b & \sigma_b\lambda_b \\
-g_a/\varepsilon_b & 0 & 0 & 0 & 0 & 0 & 1/\varepsilon_b & -c_b/\varepsilon_b
\end{pmatrix}$$

Here subscripts ($a, b$) denotes the TF genes A and B respectively. Various matrix elements are defined as follows.

$$A_h = \mu_h \left( \kappa_k + \lambda_k \right) \left( 1 - \beta_h \eta_h \right) n_k / \eta_k \; ; \; g_h = \mu_h / \beta_h \eta_h \; ; \; c_h = \kappa_h + \lambda_h + \chi_h A_k$$

In these definitions for $h = a, b$ one needs to substitute $k = b, a$. To obtain the critical value of the Hill coefficient ($_c n_a$) one need to first solve the steady state equations. Upon substituting the steady state protein values into to the corresponding Jacobian matrix one can construct the characteristic polynomial and the Routh table. The corresponding inequality conditions for oscillations will be derived from this table. This procedure need to be carried out at various values of $n_a$ from $n_a = 1$ in an iterative way. We set the default initial value of promoter state variable as $X_h = 5 \times 10^{-2}$ for any one of the TF genes A/B/C to trigger the limit-cycle oscillations in case of repressilator so that the effect of other Group I control parameters on the initial delay in oscillations can be studied. One should note that this initial delay is still a function of the disproportion in the initial conditions. We measure the concentrations of various molecules in terms of number of molecules inside the cell. Considering a typical *E. coli* bacterial cell (volume $\sim 10^{-18}\,\text{m}^3$) [39] we set $d_{hz} = 1$, $m_{hs} \sim 10^2$ molecules and $p_{hs} \sim 10^4$ molecules [8], [9] where $h = a, b$ and $c$ respectively denotes TF genes A, B and C. Concentration of a single TF molecule inside a bacterial cell will be $\sim 2$ nM [40]-[42]. We assume a lifetime of TF protein as $\sim 2$ min and mRNA lifetime as $\sim 0.2$ min for *E. coli* that gives a decay rate of $\gamma_{pa} \sim 10^{-2} s^{-1}$ so that $w_a = 0.1$ and we measure the time in terms of number of protein lifetimes. Since the dynamics of binding-unbinding of a single transcription factor protein ($n_a = 1$) with its *cis*-regulatory site on DNA is a typical diffusion-controlled site-specific DNA-protein interaction, under *in vivo* conditions of a bacterial cell we find that $k_{hf} \sim 10^{-7}$ molecules$^{-1}$ ($\gamma_p$) that will be scaled down to the interaction of $n_h$ number of TF molecules with sequentially located combinatorial binding sites as $k_{hf} \Rightarrow \left( k_{hf} / n_h \right)$ molecules$^{-n_a} s^{-1}$ [26], [27]. The time associated with the unbinding reaction will be closer to that of the protein lifetime. Here we have assumed an *in vivo* 3D diffusion controlled collision rate $\sim 10^4\,\text{M}^{-1}\text{s}^{-1}$. In case of nucleus of yeast cell $k_{hf} \sim 10^{-8}$ molecules$^{-1}$ $\gamma_p^{-1}$ and in case of nucleus of human cell we find $k_{hf} \sim 10^{-9}$ molecules$^{-1}$ $\gamma_p^{-1}$. Since $v_a \sim 10^{-4}$ for a typical bacterial promoter [8], [9] we assume $\left( \gamma_{ph} / p_{as}^{n_a} k_{hf} \right) \sim 10^{-4}$ for an arbitrary Hill coefficient $n_a$. Using these values one can estimate the physiological ranges of various groups of control parameters as given in **Table 1**. With these settings for a bacterial *lacI* ($n_a = 4$) system (**Eqs. (4-5)**) the values of parameters will be $v_a < 10^{-4}$ and $\mu_a \sim 10^{-5}$. For simplification purpose we can assume identical values for similar group of parameters $n_h = n_{hq}$, $\mu_h = \mu_{hq}$ and so on for other family of parameters such as $\chi_h$ and $\varepsilon_h$.

# References

1. Lewin B (2004) Genes VIII. Oxford University Press, London.
2. Monod J, Pappenheimer AM, Cohen-Bazire G (1952) The kinetics of the biosynthesis of beta-galactosidase in Escherichia coli as a function of growth. Biochim. Biophys. Acta. 9:648-660.
3. Ptashne M, Gann A (2002) Genes and Signals. Cold Spring Harbor Laboratory Press, New York.
4. Wagner R (2000) Transcription Regulation in Prokaryotes. Oxford University Press, Oxford.
5. Alon U (2006) An Introduction to Systems Biology. CRC Press, London.
6. Francois P, Hakim V (2005) Core genetic module: The mixed feedback loop. Physical Reviews E 72:031908.
7. Amir A, Kobiler O, Rokney A, Oppenheim A, Stavans J (2007) Noise in timing and precision of gene activities in a genetic cascade. Mol Syst Biol 3:71-81.
8. Murugan R (2012) Theory on the dynamics of feed-forward loops in the transcription factor networks. PLoS ONE. a7(7): e41027-42.
9. Murugan R, Kreiman G (2011) On the minimization of fluctuations in the response-times of gene-regulatory networks. Biophys. J. 101:1297-1306.
10. Alon U, Surette M, Barkai N, Leibler S (1999) Robustness in bacterial chemotaxis. Nature 397:168-171.
11. Rosenfeld N, Elowitz MB, and Alon U (2002) Negative autoregulation speeds the response times of transcription networks. J Mol Biol 323:785-793.
12. Shen-Orr S, Milo R, Mangan S, Alon U (2002) Network motifs in the transcriptional regulation network of Escherichia coli. Nature Genetics 31:64-68.
13. Pomerening JR, Kim SY, Ferrell JE (2005) Systems-level dissection of the cell-cycle oscillator: Bypassing positive feedback produces damped oscillations. Cell 122: 565–578.
14. Goldbeter A (2002) Computational approaches to cellular rhythms. Nature 420: 238–245.
15. Forger DB, Peskin CS (2005) Stochastic simulations of the mammalian circadian clock. Proc Natl Acad Sci USA 102: 321–324.
16. Goodwin BC (1965) Oscillatory behavior in enzymatic control processes. Adv. Enzyme Reg. 3:425–439.
17. Griffith JS (1968) Mathematics of cellular control processes I. Negative feedback to one gene J. Theor. Biol. 20:202-208.
18. Purcell O, Savery NJ, Grierson CS, Bernardo MD (2010) A comparative analysis of synthetic genetic oscillators. J. R. Soc. Interface 7:1503-24.
19. Tyson JJ, Albert R, Goldbeter A, Ruoff P, Sible J (2008) Biological switches and clocks J. R. Soc. Interface 5: S1-S8.
20. Bratsun D, Volfson D, Tsimring LS, Hasty J (2005) Delay-induced stochastic oscillations in gene regulation. Proc. Natl Acad. Sci. USA 102:14 593–14 598.
21. Atkinson MR, Savageau MA, Myers JT, Ninfa AJ (2003) Development of genetic circuitry exhibiting toggle switch or oscillatory behavior in Escherichia coli. Cell 113: 597–607.
22. Guantes R, Poyatos JF (2006) Dynamical principles of two-component genetic oscillators. PLoS Comput. Biol. 2:e30.
23. Lewis J (2003) Autoinhibition with transcriptional delay: A simple mechanism for the zebrafish somitogenesis oscillator. Curr. Biol 13: 1398-1408.

24. Elowitz MB, Leibler S (2000) A synthetic oscillatory network of transcriptional regulators. Nature 403: 335–338.
25. Stricker J, Cookson S, Bennett MR, Mather WH, Tsimring LS, Hasty J (2008) A fast, robust and tunable synthetic gene oscillator. Nature 456:516–519.
26. Murugan R (2010) Theory of site-specific interactions of the combinatorial transcription factors with DNA. J. Phys. A: Math. & Theor. 43:195003–23.
27. Murugan R (2010) Theory on the mechanism of distal-action of transcription factors: Looping of DNA versus tracking along DNA. J. Phys. A: Math. & Theor. 43:415002–17.
28. Murugan R (2010) Theory of site-specific DNA-protein interactions in the presence of conformational fluctuations of DNA binding domains. Biophys. J. 99:353-359.
29. Murugan R (2007) Generalized theory of site-specific DNA-protein interactions. Physical Review E 76:011901-10.
30. Routh EJ (1877) A treatise on the stability of motion, Macmillan & Co (London).
31. Strogatz SH (2000) Nonlinear dynamics and chaos: With applications in physics, biology, chemistry and engineering. Massachusetts: Perseus Publishing. 498 p.
32. Fall CP, Marland ES, Wagner JM, Tyson JJ (2002) Computational cell biology. New York: Springer-Verlag. 488 p.
33. Smolen P, Baxter DA, Byrne JH (2002) A reduced model clarifies the role of feedback loops and time delays in the Drosophila circadian oscillator. Biophys J. 83:2349-59.
34. Fung E, Wong WW, Suen JK, Bulter T, Gu Lee S, Liao JC (2005) A synthetic gene-metabolic oscillator. Nature 435:118–122
35. Tsai TY, Choi1 YS, Ma W, Pomerening JR, Tang C, Ferrell JE (2008) Robust, Tunable Biological Oscillations from Interlinked Positive and Negative Feedback Loops. Science 321: 126-129.
36. Shahrezaei V, Ollivier J, Swain P (2008) Colored extrinsic fluctuations and stochastic gene expression. Mol Syst Biol 4:196-205.
37. Shahrezaei V, Swain PS (2008) Analytical distributions for stochastic gene expression. Proc Natl Acad Sci USA 105:17256-17261.
38. Press WH, Teukolsky S, Flannery BP, Vetterling WT (2007) Numerical Recipes in C: The Art of Scientific Computing. Cambridge University Press.
39. Goodsell DS (1991) Inside a living cell. Biochem Sci 16**:** 203-206.
40. Neumann FR, Nurse P (2007) Nuclear size control in fission yeast. J Cell Biol 179: 593-600.
41. Albe KR, Butler MH, Wright BE (1990) Cellular concentrations of enzymes and their substrates. Journal of Theoretical Biology 143:163-195.
42. Newman J, et. al. (2006) Single-cell proteomic analysis of S. cerevisiae reveals thearchitecture of biological noise. Nature 441:840-846.

## FIGURES CAPTIONS

**Figure 1**
Goodwin-Griffith genetic oscillator model. The transcription factor (TF) gene A is transcribed and translated into the TF protein product that in turn is converted to the active end-product. The end-product (or its oligomer as in case of *lacI* repressor negative-feedback-only system that was constructed in **Ref. 25**) is the key molecule that locates the respective *cis*-regulatory elements associated with the promoter of TF gene A through a combination of one-dimensional (1D) and three-dimensional (3D) routes as that of typical site-specific DNA-protein interactions. Here either monomers of the end-product directly assemble at the corresponding regulatory elements in a combinatorial manner (I) or the fully-formed complex of $n_a$-mer binds with the respective regulatory sites (II). Assembly of combinatorial TF molecules results in the looping of DNA segment that is present in between the promoter and *cis*-regulatory elements. ARPC is the assembled repressor-promoter complex that in turn results in down-regulation. Our analysis shows that out of these two competing pathways, the pathway I is the most probable one since it takes shortest time. The corresponding set of differential equations is given in **Eqs (4-5)**. This system is well characterized by parameters of Group I, II and III. Group I consists of parameters $\left(w_a, v_a, \varepsilon_a\right)$ whereas Group II consists of equilibrium parameters $\left(\lambda_a, \mu_a\right)$ and Group III consists of ordinary type perturbation parameters $\left(\sigma_a, \kappa_a, \chi_a\right)$. Most of the earlier studies assumed zero values to Group II parameters apart from assuming zero for $v_a$ that controls the promoter state dynamics. Blue colored spheres are the dimers of *lac* repressor. Here a.a denotes amino acids and n.a denotes nucleotides.

**Figure 2**
In case of positive regulation the combinatorial transcription factors bound at *cis*-regulatory modules enhance the initiation of transcription by strengthening the RNAPII-promoter interactions through their distal action (positive arrows) whereas in case of negative regulation, the RNAPII-promoter complex will be destabilized by the combinatorial TFs present at CRMs (negative arrows).
**A.** Goodwin-Griffith oscillator.
**B1**. One-to-one dual feedback oscillator. Here the end-product of TF gene A binds at the promoter of TF gene B and down-regulates it whereas the end-product of TF gene B binds with the promoter of TF gene A and down-regulates it.
**B2**. Two independent Goodwin-Griffith oscillators are coupled through -OR- type logic with NN-NN configuration. Here the promoter of TF gene A will have binding sites for the end-products of both TF gene A and B and so on for TF gene B.
**B3**. GG oscillators are coupled through -AND- type logic with N-N configuration.
**B4**. GG oscillators are coupled through -OR- type logic with NN-PP configuration.
**B5**. GG oscillators are coupled through -OR- type logic with NP-NP configuration.
**B6.** Possible robust synthetic gene oscillator. Here K is the booster TF gene that is coupled to N-N type dual feedback oscillator via –OR- gate.
**C1.** Repressilator that is built with three TF genes by cyclic coupling.
**C2**. Three independent Goodwin-Griffith modules are coupled through -OR-type logic. Here dashed lines show the fully interconnected network.

**C3**. Three independent Goodwin-Griffith modules with -AND-type logic. Here dashed lines show the fully interconnected network.

**Figure 3**

**A1**. Phase portraits of Goodwin-Griffith oscillator as described by **Eqs 4**. One needs to substitute Q = M (scaled concentration of mRNA) for red line, Q = X (promoter occupancy) for blue line and Q = Z (end-product) for pink line. Simulation settings are $\mu_a = 2 \times 10^{-4}$, $v_a = 10^{-3}$, $(\sigma_a, \kappa_a, \chi_a) = 0$ and we set $(\varepsilon_a, \lambda_a, w_a) = 1$ which required a critical Hill coefficient of $_cn_a = 6$ to generate oscillations. Total simulation time is 100 (measured in terms of number of lifetimes of the protein product of TF gene A) and integration step is $\Delta\tau = 10^{-5}$.

**A2**. Trajectories corresponding to the settings in A1.

**A3**. Roots of the (biquadratic) characteristic polynomial (PI) associated with the Jacobian matrix for settings in A1.

**B1**. Variation of critical Hill coefficient with the parameter set $(\mu_a, v_a)$. Minimum of this critical value seems to be achieved at $\mu_a \sim 10^{-4}$, and $v_a \sim 10^{-3}$.

**B2.** Variation of critical Hill coefficient with the parameter set $(\mu_a, \varepsilon_a)$. With the optimized settings in B1, the system seems to be robust when $\varepsilon_a \in (0.2, 2)$.

**B3.** Variation of critical Hill coefficient with the parameter set $(\mu_a, w_a)$. With the optimized settings in B1, the system seems to be robust when $w_a \in (0.2, 2)$.

**B4.** Variation of critical Hill coefficient with the parameter set $(\mu_a, \lambda_a)$. Default values of other parameters in B1-4 are as in A1.

**C1**. Variation of period and amplitude of the negative-feedback-only model considered in **Ref. 25** with respect changes in the promoter affinity parameter $\mu_a$. Simulation settings are given in **Table 1**. Red solid line in the period and blue solid line is amplitude of the oscillator.

**Figure 4**

**A1.** Phase portraits of the regular Goodwin oscillator as described by **Eqs 13**. Q = M for red line and Q = Z for pink line. Simulation settings are $(\sigma_a, \kappa_a) = 0$, $\mu_a \sim 10^{-12}$ and $(\varepsilon_a, \lambda_a, w_a) = 1$ which require a critical Hill coefficient of $_cn_a = 9$. Total simulation time is 100 (measured in terms of number of lifetimes of the protein product of TF gene A) and integration scaled-time step is $\Delta\tau = 10^{-5}$.

**A2**. Trajectories corresponding to the settings in A1.

**A3**. Roots of the (cubic) characteristic polynomial (PII) associated with the Jacobian matrix for settings in A1.

**B1.** Variation of critical Hill coefficient with the parameter set $(\mu_a, w_a)$. With the optimized settings in B1, the system seems to be robust when $w_a \in (0.3, 2.5)$.

**B2.** Variation of critical Hill coefficient with the parameter set $(\mu_a, \varepsilon_a)$. With the optimized settings in B1, the system seems to be robust when $\varepsilon_a \in (0.5, 1.5)$.

**B3.** Variation of critical Hill coefficient with the parameter set $(\mu_a, \lambda_a)$.

**B4.** Variation of critical Hill coefficient with the parameter set $\left(\kappa_a, \sigma_a\right)$. Default values of other parameters in B1-4 are as in A1.

## Figure 5

**A1.** Effects of perturbations in $v_a$ on the limit-cycle orbit of GG oscillator. The default simulation settings are $\mu_a = 2 \times 10^{-5}$ and $v_a = 5 \times 10^{-5}$, $\left(\sigma_a, \kappa_a, \chi_a\right) = 0$, $\left(\varepsilon_a, \lambda_a, w_a\right) = 1$ which required a critical Hill coefficient of $_c n_a = 6$. Perturbation introduced in the interval from time 30 to 100 by abruptly raising the value to $v_a = 10 \times 10^{-5}$. Increase in $v_a$ increases the period and reduce the amplitude of oscillations. Total simulation time is 200 (measured in terms of number of lifetimes of the protein product of TF gene A) and integration step is $\Delta \tau = 10^{-5}$.

**A2.** Effects of perturbations in $w_a$. The default simulation settings are as in A1. Perturbation introduced in the interval from time 30 to 100 by abruptly raising the value to $w_a = 2$. Increase in $w_a$ increases the period and reduce the amplitude.

**A3.** Effects of perturbations in $\varepsilon_a$. The default simulation settings are as in A1. Perturbation introduced in the interval from time 30 to 100 by abruptly raising the value to $\varepsilon_a = 2$. Increase in $w_a$ increases the period and reduce the amplitude.

**B1**. Effects of variation of $w_a$ on the period and critical Hill coefficient that is required to generate oscillations. With the current default settings, there exists a range of $w_a \in \left(0.5, 2\right)$ at which the critical Hill coefficient is minimum.

**B2**. Effects of variation of $\sigma_a$ on the period and critical Hill coefficient. Both period of oscillations and critical Hill coefficient are linearly dependent on $\sigma_a$ and one cannot assume $\sigma_a = 0$ as in cases of several earlier studies.

## Figure 6

**A1-2**. Phase portraits of TF genes A and B which are coupled through one-to-one cross feedback loops. Simulation settings are $\left(\sigma_h, \kappa_h, \chi_h\right) = 0$, $\left(\varepsilon_h, \lambda_h, w_h, \rho_h\right) = 1$, $\mu_h = 10^{-5}$ and $v_h = 5 \times 10^{-5}$ which required a critical Hill coefficient of $_c n_a = 6$. Total simulation time is 100 (measured in terms of number of lifetimes of the protein product of TF gene A) and integration step is $\Delta \tau = 10^{-5}$.

**A3-4**. Trajectories of TF genes A and B. Perturbation in $\mu_a$ was introduced at $\tau_{pulse} = 20$ by abruptly raising $\mu_a$ to $\mu_a + 10^{-8}$ for a period of $\tau_w = 10^{-3}$. This corresponds to a disproportion of $\pi_\mu = 10^{-1}$ (%). The system quits the limit-cycle orbit with a delay of $\tau_{del} \approx 50$. Other default simulations settings are as in A1-2. Q = X for red line, Q = M for green line and Q = P for blue line.

**A5**. Phase portraits of TF genes A and B. Q = P for red line, Q = M for blue line and Q = X for green line.

**A6**. Roots of the eighth dimensional characteristic polynomial derived from the Jacobian matrix associated with **Eqns 18** for the parameter settings given in A1-2.

**B1-8**. Variation of critical Hill coefficient that is required to generate oscillations with respect to changes in various control parameters. Default settings of other parameters are as in A1-2.

**C**. Variation of $\tau_{del}$ with respect changes in percentage of disproportion $\pi_\mu$ and pulse width $\tau_w$.

**Figure 7**

**A1-2.** Trajectories of protein-products of TF genes A and B which are two independent GG oscillators coupled through A-OR-B type logic as given in **Figure 2B2**. Simulation settings are $\left(\sigma_h, \kappa_h, \chi_{hq}\right) = 0$, $\left(\varepsilon_h, \lambda_h, w_h, \rho_h\right) = 1$, $\mu_h = 10^{-5}$ and $v_h = 10^{-5}$ which required a critical Hill coefficient of $_c n_a = 8$. Total simulation time is 200 (number of lifetimes of the protein product of TF gene A) and integration step is $\Delta\tau = 10^{-5}$. For each promoter A/B the parameter $v_h$ will be split into $v_h \Rightarrow \left(v_{hh}, v_{hq}\right)$ where $v_{hh}$ corresponds to self-regulation and $v_{hq}$ corresponds to cross regulation. Under identical values of all the parameters the system generates synchronized oscillations with a period of $\tau_p \sim 4.5$. Upon introduction of perturbation in $v_a$ from scaled time 0 to 100 (where $v_a = 15 \times 10^{-5}$), the amplitude of TF gene A is reduced with a phase shift and the period of entire system that includes both TF genes A and B remains the same. A2 is a magnification of certain range of A1.

**A3-4.** Effect of perturbation in only one of the split parameters $\left(v_{hh}, v_{hq}\right)$ associated with TF gens A/B. Here $v_{aa}$ is perturbed to $v_{aa} = 15 \times 10^{-5}$. The system seems to be unstable and generates beats. A4 is a magnification of certain range of A3.

**A5-6.** Here $v_{aa}$ is perturbed to $v_{aa} = 30 \times 10^{-5}$. Period of beats seems to increase as the disproportion among the split parameters increases. A6 is a magnification of certain range of A5.

**B1-2.** Effect of perturbation in the parameter $w_a$ which is raised to $w_a = 1.5$ from the default value in the interval from 0 to 100. Increase in $w_a$ reduces the amplitude of both the TF genes A and B with a phase shift and the period of oscillations of the entire system remains the same as in A1-2. B2 is a magnification of certain range of B1.

**C1-2.** Effect of perturbations in the parameter $\varepsilon_a$ which is raised to $\varepsilon_a = 1.2$ from the default value in the time interval from 0 to 100. Increase in $\varepsilon_a$ reduces the amplitude of both the TF genes A and B without a phase shift and the period of oscillations of the entire system remains the same as in A1-2. C2 is a magnification of certain range of C1.

**Figure 8**

**A1-4.** Trajectories of protein-products of TF genes A, B which are two independent GG oscillators coupled through A-AND-B type logic as given in **Figure 2B3**. Simulation settings are $\left(\sigma_h, \kappa_h, \chi_{hq}\right) = 0$, $\left(\varepsilon_h, \lambda_h, w_h, \phi_d, \rho_h\right) = 1$, $\mu_h = 10^{-5}$ and $v_h = 10^{-5}$ which required $_c n_a = 2$ (here we have set $n_h = 4$ for clarity of results). Total simulation time is 200 (number of lifetimes of the protein product of TF gene A) and integration step is $\Delta\tau = 10^{-5}$. Under identical values of all the parameters the system generates synchronized oscillations with a period of $\tau_p \sim 9$. Upon introduction of perturbation in $v_a$ from scaled time 0 to 100 (where $v_a = 15 \times 10^{-5}$ in A1, $v_a = 30 \times 10^{-5}$ in A3), the amplitude of TF gene A is reduced with a phase shift and the period of entire system (both TF genes A and B) increases to $\tau_p \sim 10$ in A1. A2 and A4 are magnifications of certain range of A1 and A3.

**B1-2**. Effect of perturbation in the parameter $w_a$ which is raised to $w_a = 2$ from the default value in the time interval from 0 to 100. Increase in $w_a$ increases the period of oscillations of the entire system to $\tau_p \sim 9.5$ as in A1-2 and reduces the amplitude of TF genes A with a phase shift. B2 is a magnification of certain range of B1.

**C1-2**. Effect of perturbations in the parameter $\varepsilon_a$ which is raised to $\varepsilon_a = 10$ from the default value in the time interval from 0 to 100. Increase in $\varepsilon_a$ increases the period of oscillations of the entire system to $\tau_p \sim 9.5$ as in A1-2 and reduces the amplitude of both TF genes A and B without a phase shift. C2 is a magnification of certain range of C1.

**Figure 9**

**A1-4**. Phase portraits and trajectories of TF genes A, B and C of a repressilator. Simulation settings are $\left(\sigma_h, \kappa_h, \chi_h\right) = 0$, $\left(\varepsilon_h, \lambda_h, w_h, \rho_h\right) = 1$, $\mu_h = 10^{-4}$ and $v_h = 10^{-4}$ which required a critical Hill coefficient of $_cn_a = 2$. Total simulation time is 200 (number of lifetimes of the protein product of TF gene A) and integration step is $\Delta\tau = 10^{-5}$. To trigger the oscillations, we have introduced the asymmetry in the initial condition for the promoter state occupancy of TF gene A as $X_a = 5 \times 10^{-2}$. Oscillations starts with a time delay $\tau_{del}$ whose value depends on the magnitude of this disproportion in the parameter values.

**A5**. Roots of the twelfth degree characteristic polynomial associated with the Jacobian matrix of **Eqns 26** for settings given in A1.

**B1-2**. Effects of perturbation in $v_a$ that is raised to $v_a = 10^{-3}$ ($v_a = 10^{-2}$ in B2) in the time interval from 0 to 100. Increase in $v_a$ increases the period of oscillations of the entire system from $\tau_p \sim 23$ to 24.5 and reduces the amplitudes of TF genes A and C. The amplitudes of TF genes A/B/C are such that A < C < B.

**B3-4**. Effects of perturbation in $\left(w_a, \varepsilon_a\right)$ which are raised to $w_a = 3$ in the time interval from 0 to 100. Increase in $w_a$ increases the period of oscillation of the entire system from $\tau_p \sim 23$ to 30 and reduces the amplitudes of TF genes A and B and increases the amplitude of C and the amplitudes of TF genes are such that B < A < C (**B3**). Increase in $\varepsilon_a$ increases the period of oscillation of the entire system as in B3 where the amplitudes of TF genes A/B/C are such that B < A < C (**B4**).

**Figure 10**

**A1-3**. Phase portraits of TF genes A/B/C which *are* three independent GG oscillators cyclically coupled through -OR- type logic as given in **Figure 2C2** (without dashed lines). Simulation settings are $\left(\sigma_h, \kappa_h, \chi_h\right) = 0$, $\left(\varepsilon_h, \lambda_h, w_h, \rho_h\right) = 1$, $\mu_h = 10^{-5}$ and $v_h = 10^{-4}$ which required a critical Hill coefficient of $_cn_a = 5$ (we have set this to 6 for clarity of results). Total simulation time is 500 (number of lifetimes of the protein product of TF gene A) and integration step is $\Delta\tau = 10^{-5}$. In this configuration the end-products of TF gene A and C will regulate TF genes A through A-OR-C type logic whereas the promoter of TF gene B will be regulated by the end-products of TF genes A and B through A-OR-B type logic and so on. Under identical values of all the parameters the system generates synchronized oscillations with a period of $\tau_p \sim 7$.

**B1-2.** Effects of perturbation in $v_h$. Upon introduction of perturbation in $v_{aa}$ from scaled time 100 to 400 (where $v_a = 6 \times 10^{-4}$) there three phases of responses. In the first phase, the system tries to resist the perturbation whereas the second phase consists of repeating elements of resistance and chaos. Upon removal of perturbation the system enters into new limit-cycle orbit in the third phase. In B2 both $v_{aa}$ and $v_{ac}$ are perturbed as in B1.

**B3-4.** Effects of perturbation in $(w_a, \varepsilon_a)$ which are raised to $(\varepsilon_a, w_a) = 2$ in the time interval from 100 to 400. System responds to the perturbation as in B1-2.

**C1-2**. Trajectories of TF genes A/B/C which are three independent GG oscillators cyclically coupled through -AND- type logic as given in **Figure 2C3** (without dashed lines) and their responses to perturbations in Group I control parameters. Simulation settings are $(\sigma_h, \kappa_h, \chi_h) = 0$, $(\varepsilon_h, \lambda_h, w_h, \phi_d, \rho_h) = 1$, $\mu_h = 10^{-5}$ and $v_h = 10^{-4}$ which required $_cn_a = 2$. Total simulation time is 500 and integration step is $\Delta\tau = 10^{-5}$. Identical values of all the parameters of the system generate synchronized oscillations. Introduction of perturbation in $v_a$ from scaled time 100 to 400 (where $v_a = 3 \times 10^{-4}$, **C1**) affects only TF gene A whereas the orbit of other TF genes B/C seems to be stable and upon removal of the perturbation the system returns back to initial limit-cycle orbit. In **C2** the parameter $w_a$ is perturbed to $w_a = 2$ as in C1.

## Figure 11

**A1-3**. Phase portraits of TF genes A/B/C which are three independent GG oscillators which are fully interconnected with -OR- type logic as given in **Figure 2C2** (with dashed lines). Simulation settings are $(\sigma_h, \kappa_h, \chi_h) = 0$, $(\varepsilon_h, \lambda_h, w_h, \rho_h) = 1$, $\mu_h = 10^{-5}$ and $v_h = 10^{-4}$ which required a critical Hill coefficient of $_cn_a = 6$. Total simulation time is 500 (number of lifetimes of the protein product of TF gene A) and integration step is $\Delta\tau = 10^{-5}$. In this configuration all the end-products of TF gene A/B/C will regulate all the three TFs through A-OR-B-OR-C type logic. Identical values of all the parameters of the system generate synchronized oscillations. Perturbation in $v_a$ from scaled time 100 to 300 (where $v_a = 5 \times 10^{-5}$, **A1**) seems to make the system unstable. In **A2** $w_a$ is perturbed to $w_a = 3$ as in A1 and in **A3** $\varepsilon_a$ is perturbed to $\varepsilon_a = 3$ as in A1.

**B1-3**. Phase portraits of TF genes A/B/C which are three independent GG oscillators fully interconnected with -AND- type logic as given in **Figure 2C3** (with dashed lines). Simulation settings are $(\sigma_h, \kappa_h, \chi_{dh}) = 0$, $(\varepsilon_h, \lambda_h, w_h, \rho_h) = 1$, $\mu_h = 10^{-5}$ and $v_h = 10^{-5}$ which required a critical Hill coefficient of $_cn_a = 2$. Total simulation time is 500 (number of lifetimes of the protein product of TF gene A) and integration step is $\Delta\tau = 10^{-5}$. In this configuration all the end-products of TF gene A/B/C will be regulated by their complex. Identical values of all the parameters of the system generate synchronized oscillations. Introduction of perturbation in $v_a$ from scaled time 100 to 300 (where $v_a = 10^{-4}$ in **B1**) seems to make the system unstable. In **B2** $w_a$ is perturbed to $w_a = 3$ as in B1. In **B3** $\varepsilon_a$ is perturbed to $\varepsilon_a = 3$ as in B1.

## Figures 12

**A, C**. Tuning capability of GG oscillators coupled through A-OR-B type logic as given in **Figure 2B2**. Default Simulation settings are $(\sigma_h, \kappa_h, \chi_{hq}) = 0$, $(\varepsilon_h, \lambda_h, w_h, \rho_h) = 1$ and $v_h = 10^{-5}$.

**B, D**. Tuning capability of GG oscillators coupled through A-AND-B type logic as given in **Figure 2B3**. Default Simulation settings are $\left(\sigma_h, \kappa_h, \chi_{hq}\right) = 0$, $\left(\varepsilon_h, \lambda_h, w_h, \phi_d, \rho_h\right) = 1$ and $\nu_h = 10^{-5}$. Plots **A** and **B** show the variation of period, critical $_C n_h$ and amplitude with respect to changes in $\mu_h$ (iterated from 5 x $10^{-7}$ to $10^{-3}$ with $w_h = 1$) whereas plots **C** and **D** show the variation of these quantities with respect to changes in $w_h$ (iterated from 0.1 to 10 with $\mu_h = 10^{-5}$). Here period of oscillator is measured in the number of lifetimes of TF protein A ($1/\gamma_a$) and amplitude is measured in terms of number of $P_h/P_{hs}$.

**Figures 13**
**A, C**. Tuning capability of GG oscillators coupled through A-OR-B type logic as given in **Figure 2B2**. Default Simulation settings are $\left(\sigma_h, \kappa_h, \chi_{hq}\right) = 0$, $\left(\varepsilon_h, \lambda_h, w_h, \rho_h\right) = 1$ and $\mu_h = 10^{-5}$.
**B, D**. Tuning capability of GG oscillators coupled through A-AND-B type logic as given in **Figure 2B3**. Default Simulation settings are $\left(\sigma_h, \kappa_h, \chi_{hq}\right) = 0$, $\left(\varepsilon_h, \lambda_h, w_h, \phi_d, \rho_h\right) = 1$ and $\mu_h = 10^{-5}$. Plots **A** and **B** show the variation of period, critical $_C n_h$ and amplitude with respect to changes in $\nu_h$ (iterated from 5 x $10^{-7}$ to $10^{-4}$ with $\varepsilon_h = 1$) whereas plots **C** and **D** show the variation of these quantities with respect to changes in $\varepsilon_h$ (iterated from 0.7 to 8 with $\nu_h = 10^{-5}$). Here period of oscillator is measured in the number of lifetimes of TF protein A ($1/\gamma_a$) and amplitude is measured in terms of number of $P_h/P_{hs}$.

**Table 1**. Simulation parameters used to integrate **Eqs. (4-5)** of Goodwin-Griffith oscillator model as constructed in Ref. [**25**] using *lacI* system ($n_a = 4$) of *E. coli*.

| Parameter | Definition | Physiological values in *E. coli* | Remarks |
|---|---|---|---|
| $v_a$ | $\gamma_{pa} / p_{as}^{n_a} k_{af}$ | ~0.0001 | promoter state dynamics |
| $w_a$ | $\gamma_{pa} / \gamma_{ma}$ | ~0.5 | relative mRNA-protein lifetimes |
| $\varepsilon_a$ | $\gamma_{pa} / \lambda_{af}$ | ~1 | end-product formation dynamics |
| $\mu_a$ | $K_{arf} / p_{as}^{n_a}$ | ~0.00002 | binding of end-product at promoter |
| $\lambda_a$ | $\lambda_{ar} / \lambda_{af}$ | ~1 | end-product equilibrium dynamics |
| $\sigma_a$ | $\lambda_{af} / \gamma_{pa}$ | ~1 | connects protein and end-product formation |
| $\chi_a$ | $p_{as}^{n_a-1} d_{az} k_{af} / \lambda_{af}$ | ~1 | connects end-product and promoter state dynamics |
| $\kappa_a$ | $\gamma_{za} / \lambda_{af}$ | ~0.01 | describes end-product decay dynamics |
| $\sigma_{ya}$ | $p_{as} \lambda_{fya} / \lambda_{af}$ | ~0.1 | |
| $\varepsilon_{ya}$ | $\gamma_{pa} / \lambda_{fya} p_{as}$ | ~0.5 | |
| $\lambda_{ya}$ | $\lambda_{rya} / p_{as} \lambda_{fya}$ | ~1 | |
| $\chi_{ya}$ | $p_{as}^{n_a-1} d_{az} k_{af} / \lambda_{fya}$ | ~1 | |
| $d_{az}$ | | 1 | molecules |
| $p_{as}$ | $k_{ma} k_{pa} / \gamma_{ma} \gamma_{pa}$ | ~10000 | molecules |
| $m_{as}$ | $k_{ma} / \gamma_{ma}$ | ~100 | molecules |

**Table 2**. Various parameters associated with dual-feedback A-OR-B and A-AND-B type oscillators and their definitions.

| Parameter | Definition | Remarks |
|---|---|---|
| $v_h$ | $\gamma_{pa} / p_{qs}^{n_a} k_{hf}$ | Describes promoter state dynamics of TD gene 'h', for h = a, b, q = b, a |
| $w_h$ | $\gamma_{ph} / \gamma_{mh}$ | relative mRNA-protein lifetimes of TF gene 'h', h = a, b |
| $\varepsilon_h$ | $\gamma_{pa} / \lambda_{hf}$ | end-product formation dynamics of TF gene 'h', h = a, b |
| $\mu_h$ | $K_{hrf} / p_{hs}^{n_a}$ | binding of end-product of TF gene A/B at promoter, h = a, b |
| $\lambda_h$ | $\lambda_{hr} / \lambda_{hf}$ | end-product equilibrium dynamics, h = a, b |
| $\sigma_h$ | $\lambda_{hf} / \gamma_{ph}$ | connects protein and end-product formation, h = a, b |
| $\kappa_h$ | $\gamma_{zh} / \lambda_{hf}$ | describes end-product decay dynamics |
| $\beta_h$ | $1 + \sigma_h \kappa_h / (\lambda_h + \kappa_h)$ | |
| $\varepsilon_d$ | $\gamma_{pa} / p_{bs} \lambda_{df}$ | Describes dimerization reaction between TF proteins A and B |
| $Y_d$ | $y_d / p_{as}$ | Scaled concentration of $z_a$-$z_b$ dimer. |
| $\chi_{dh}$ | $\lambda_{df} p_{qs} / \lambda_{hf}$ | |
| $\phi_d$ | $\lambda_{dr} / \lambda_{df} p_{bs}$ | Described dimerization equilibrium |
| $\chi_h$ | $p_{hs}^{n_h - 1} d_{hz} k_{hf} / \lambda_{hf} p_{qs}$ | |
| $\mu_{hq}$ | $K_{hqrf} / p_{hs}^{n_{hq}}$ | $\mu_h$ splits into four types of $\mu_{hq}$ in A-OR-B type coupled oscillator |
| $v_{hq}$ | $\gamma_{pa} / p_{hs}^{n_{hq}} k_{haf}$ | $v_h$ splits into four types of $v_{hq}$ in A-OR-B type coupled oscillator |
| $\chi_{hq}$ | $p_{qs}^{n_{hq}-1} d_{qz} k_{hqf} / \lambda_{hf}$ | |
| $K_{hqrf}$ | $k_{hqr} / k_{hqf}$ | Describes binding of end-product of TF gene 'q' at the promoter of TF gene 'h' |
| $\lambda_h$ | $\lambda_{hr} / \lambda_{hf}$ | Describes end-product formation equilibrium of TF gene 'h' |

Note: subscripts a denotes TF gene A and b denotes TF gene B.

Figure 1

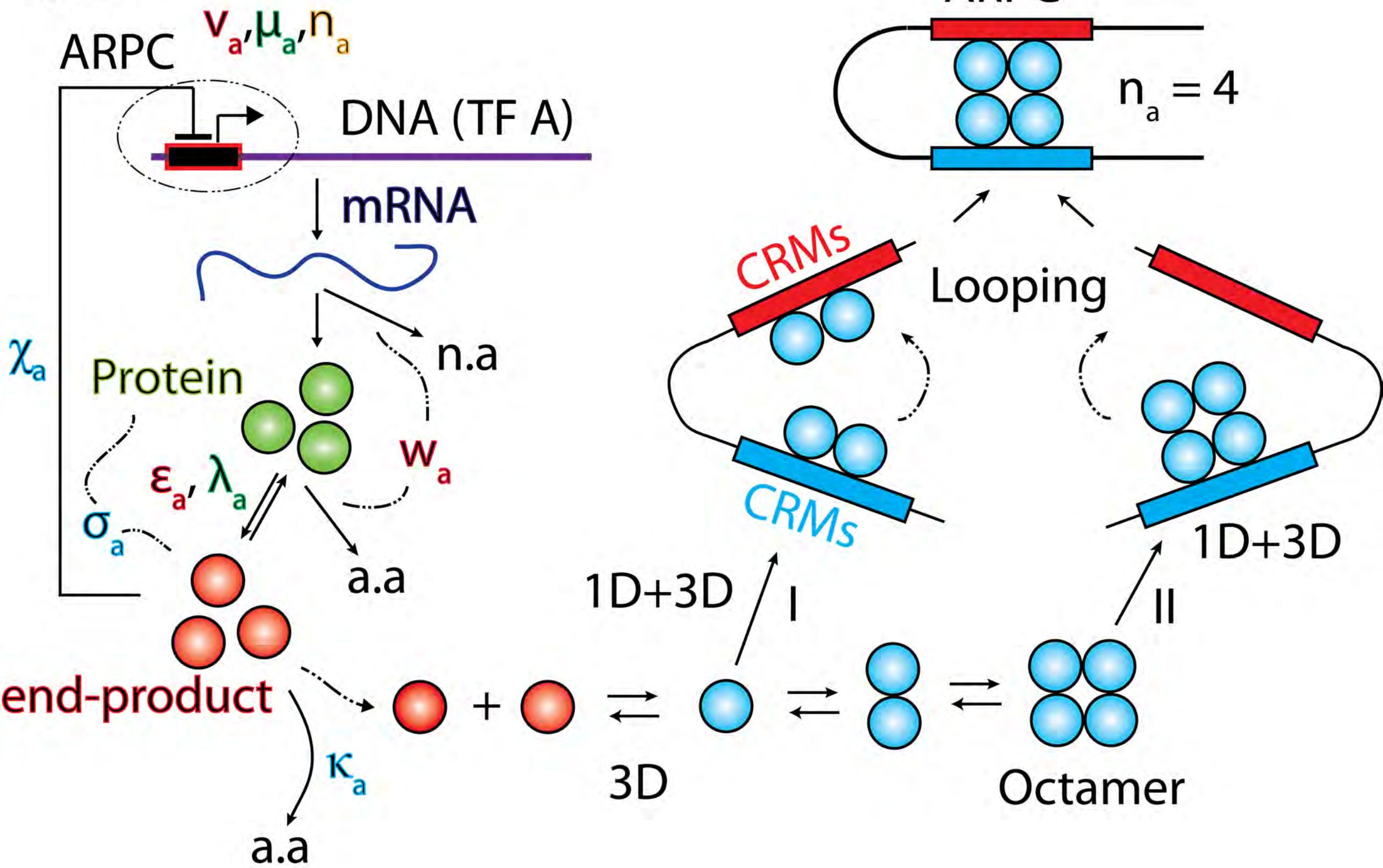

# Figure 2

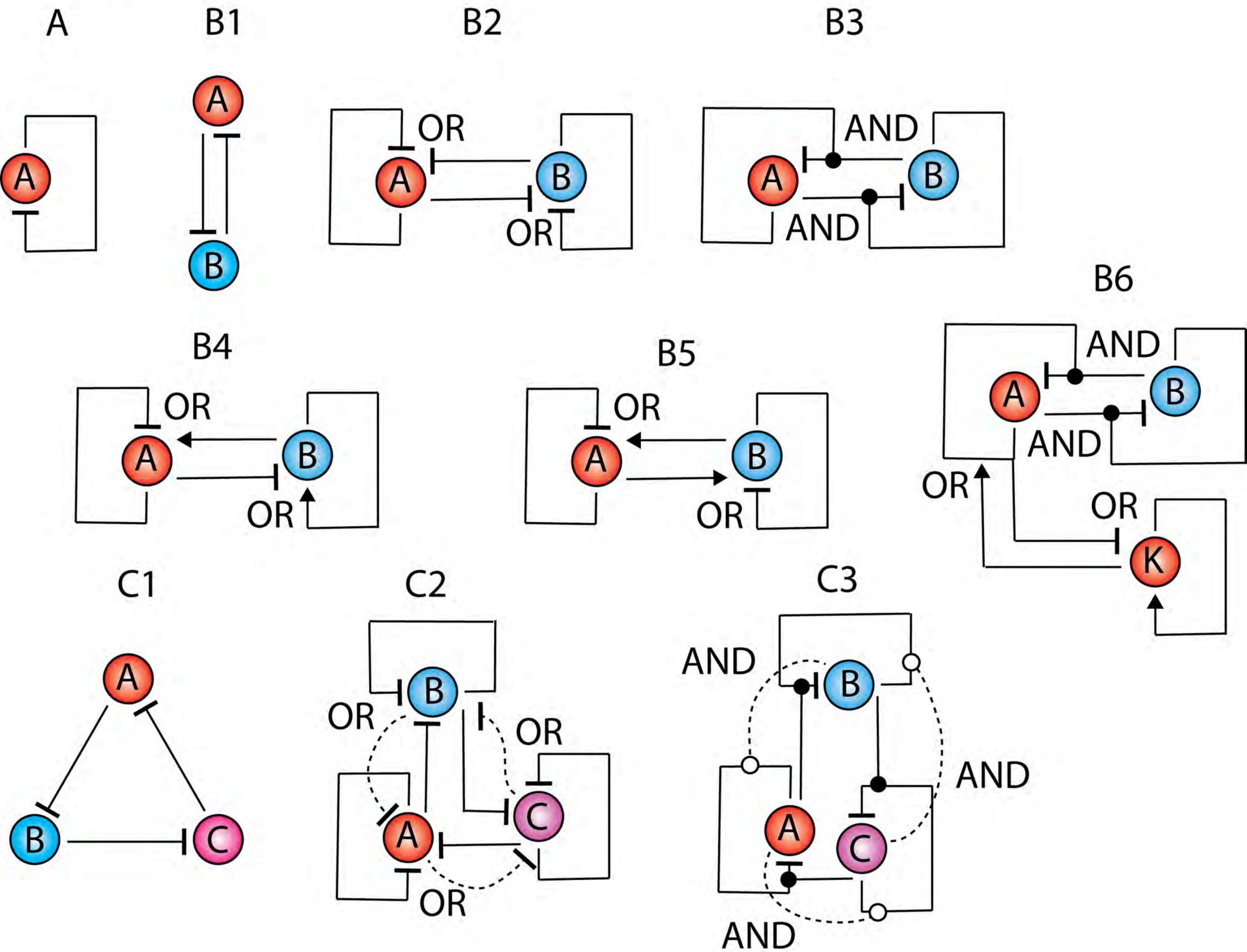

# Figure 3

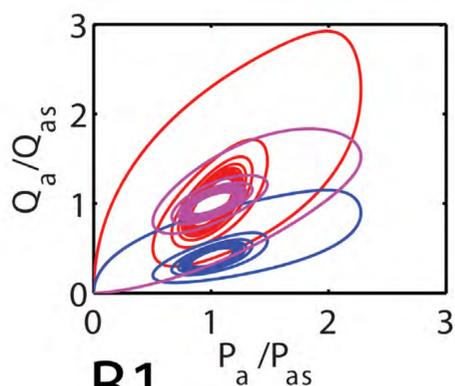

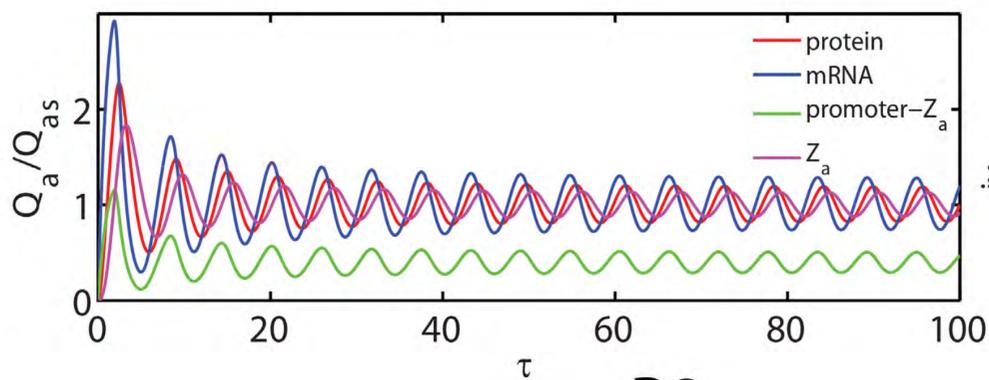

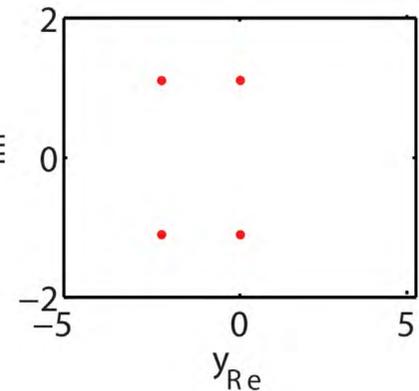

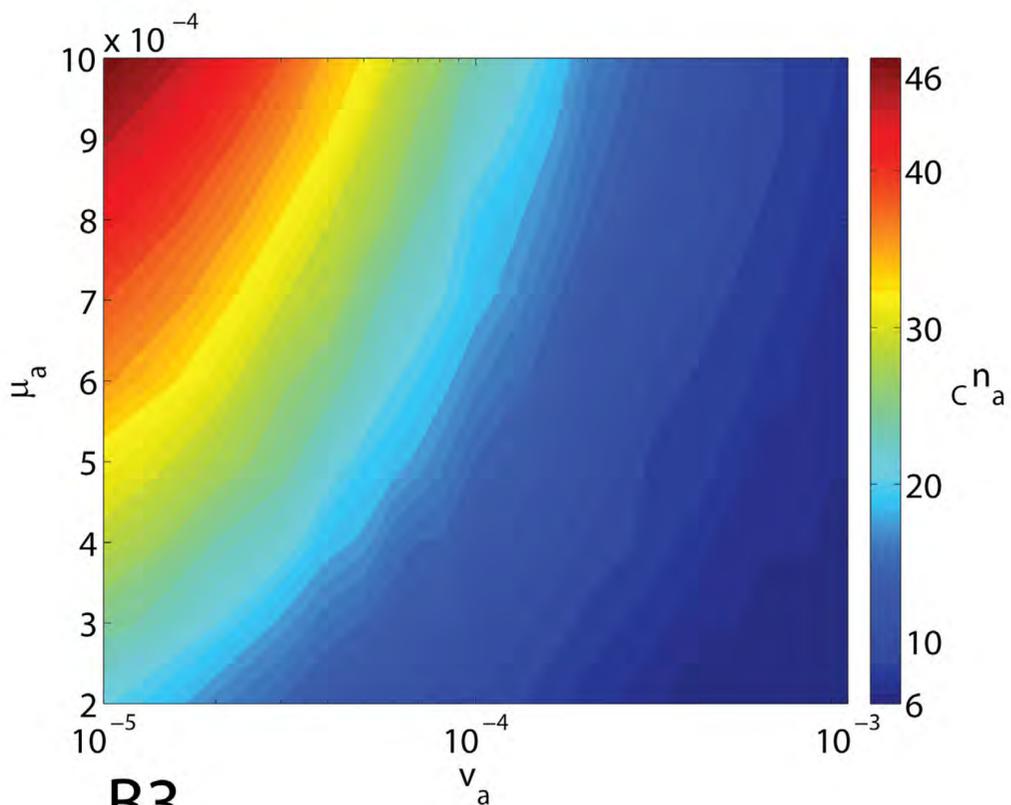

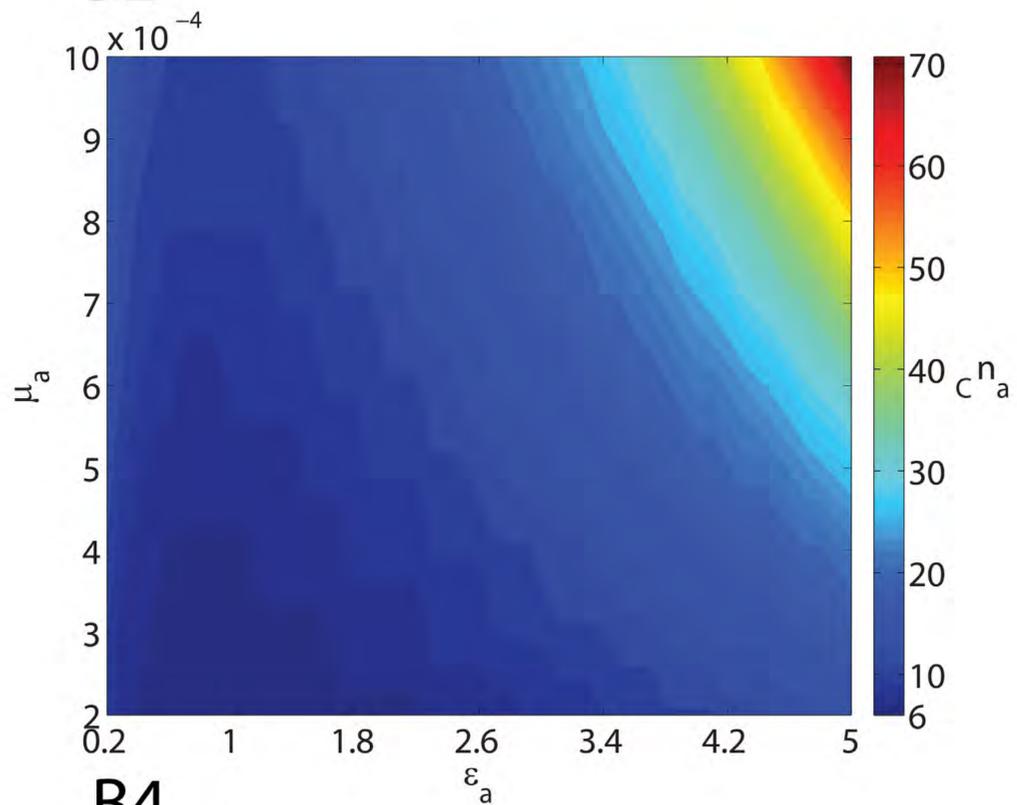

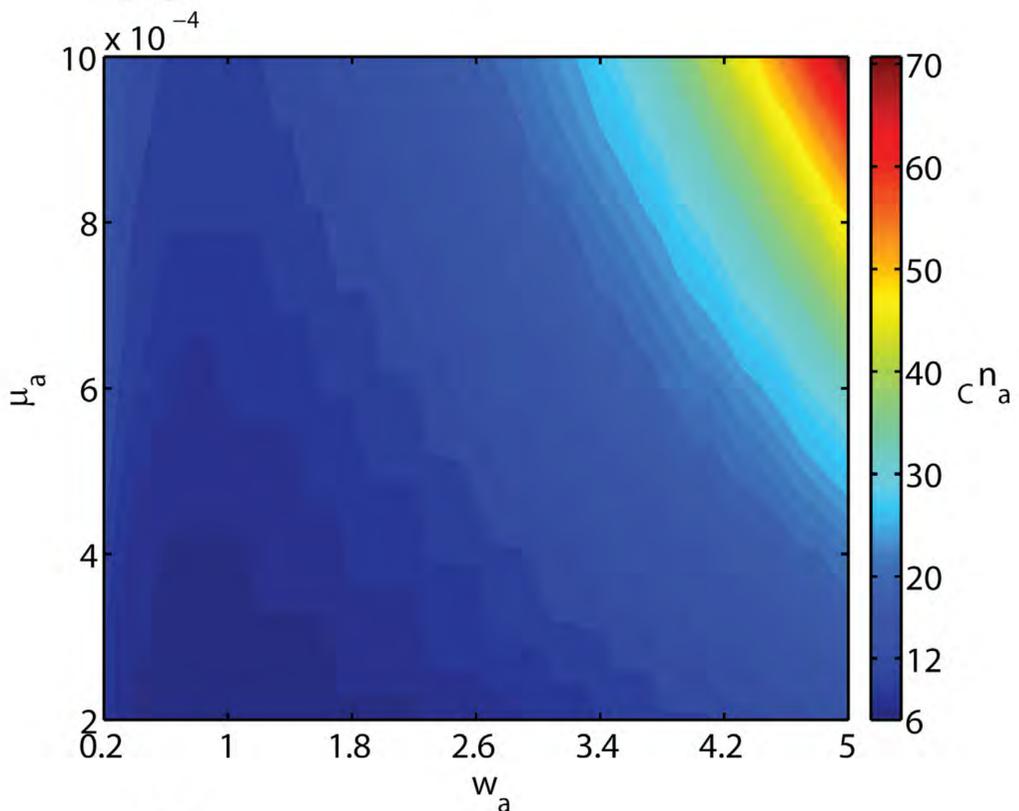

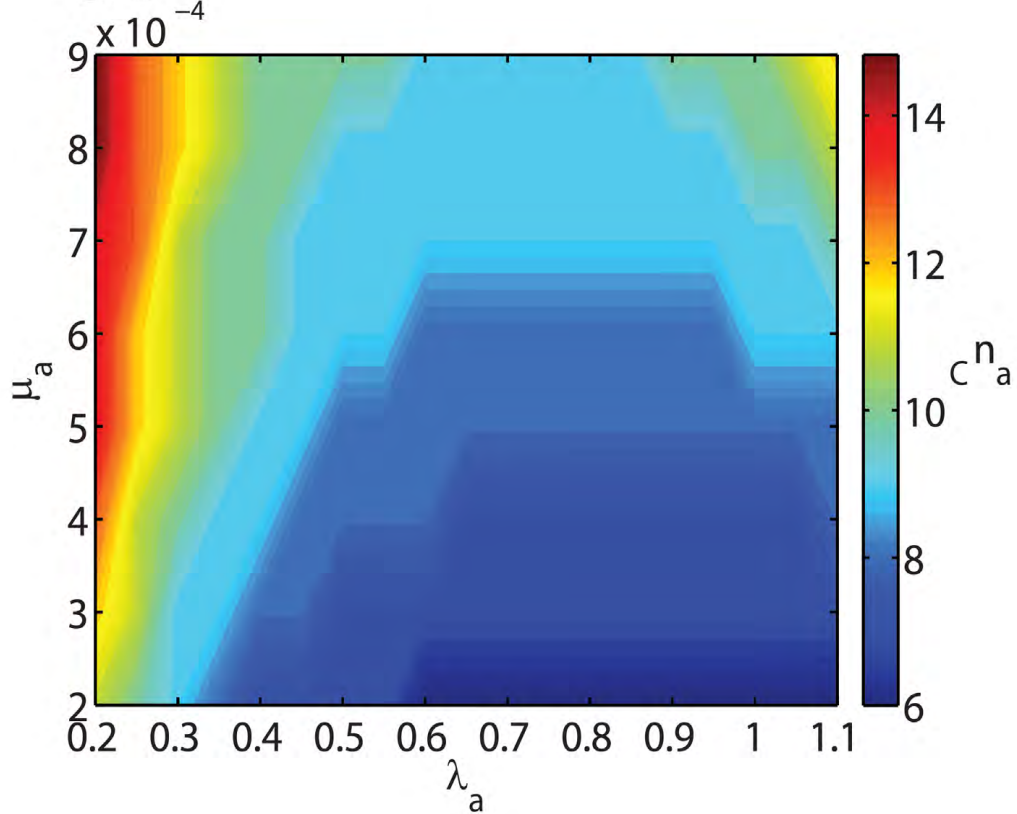

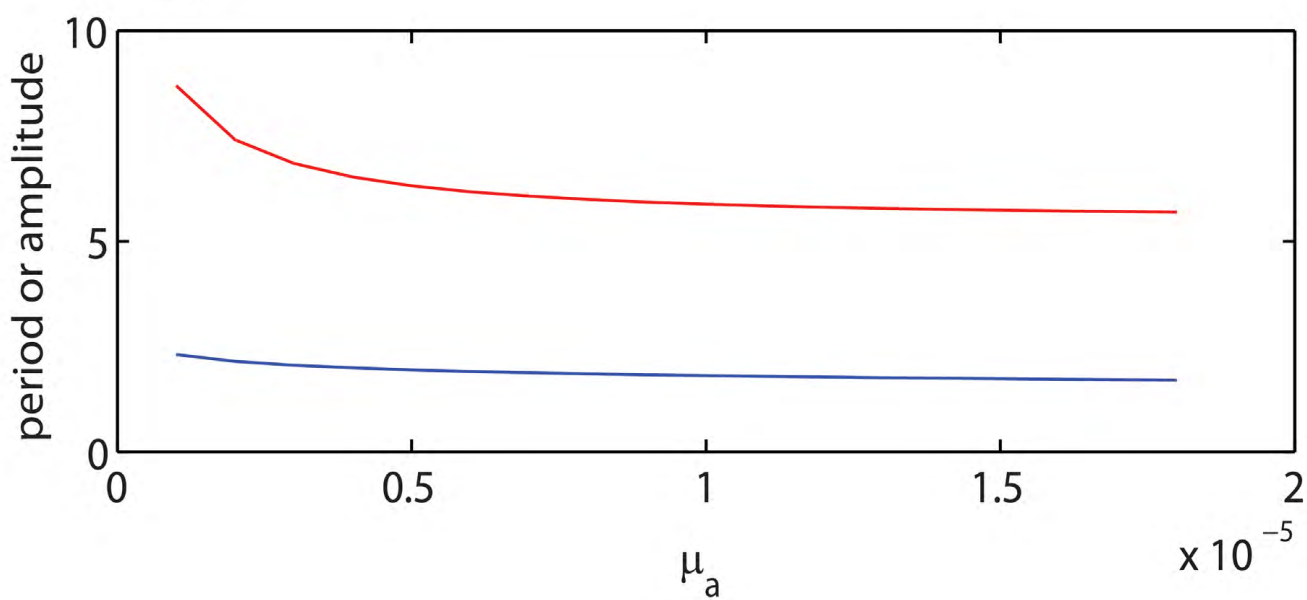

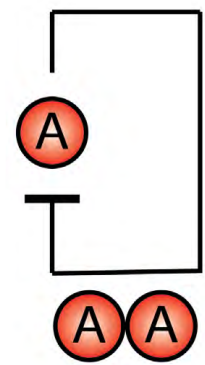

Figure 4

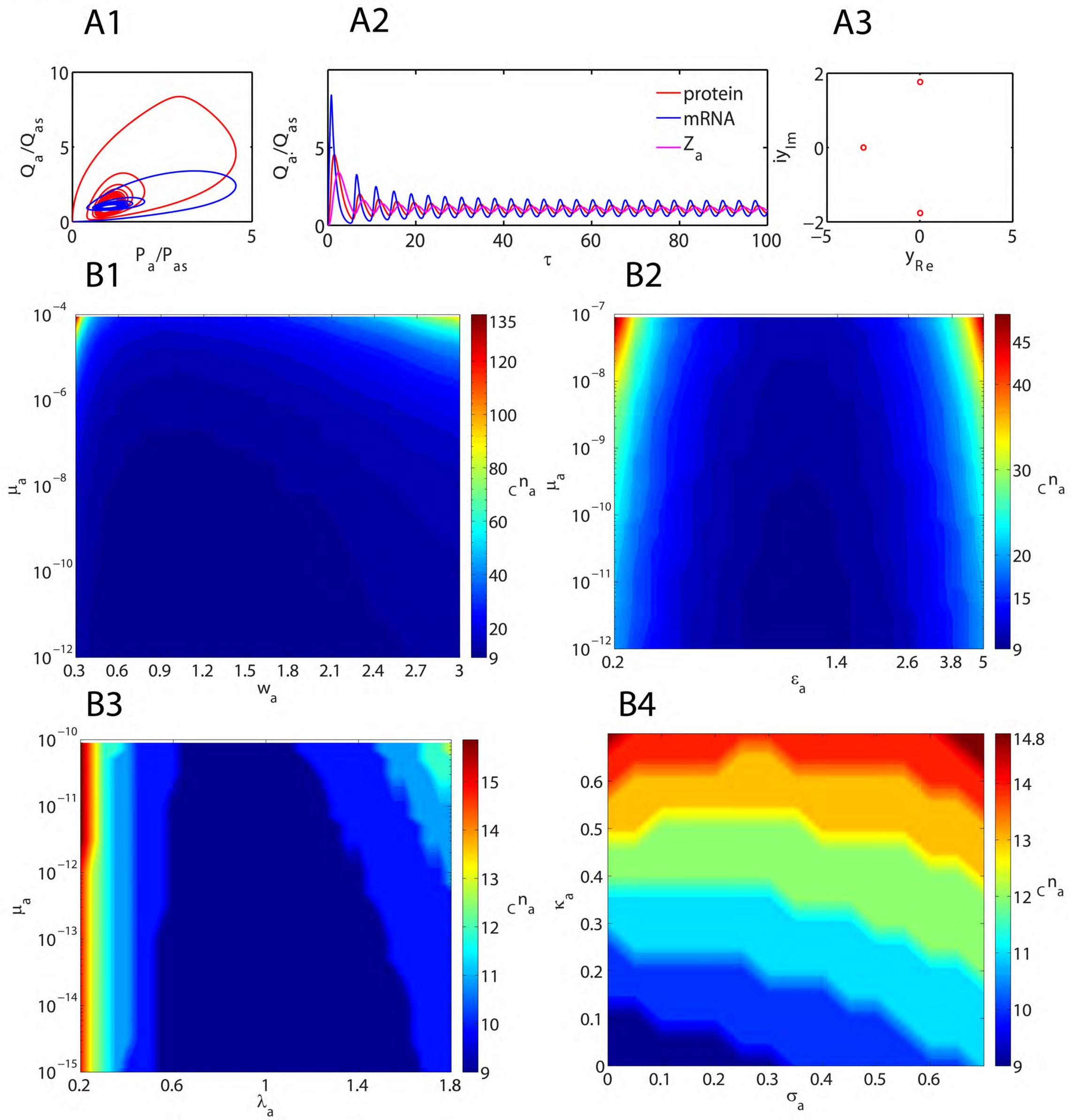

Figure 5

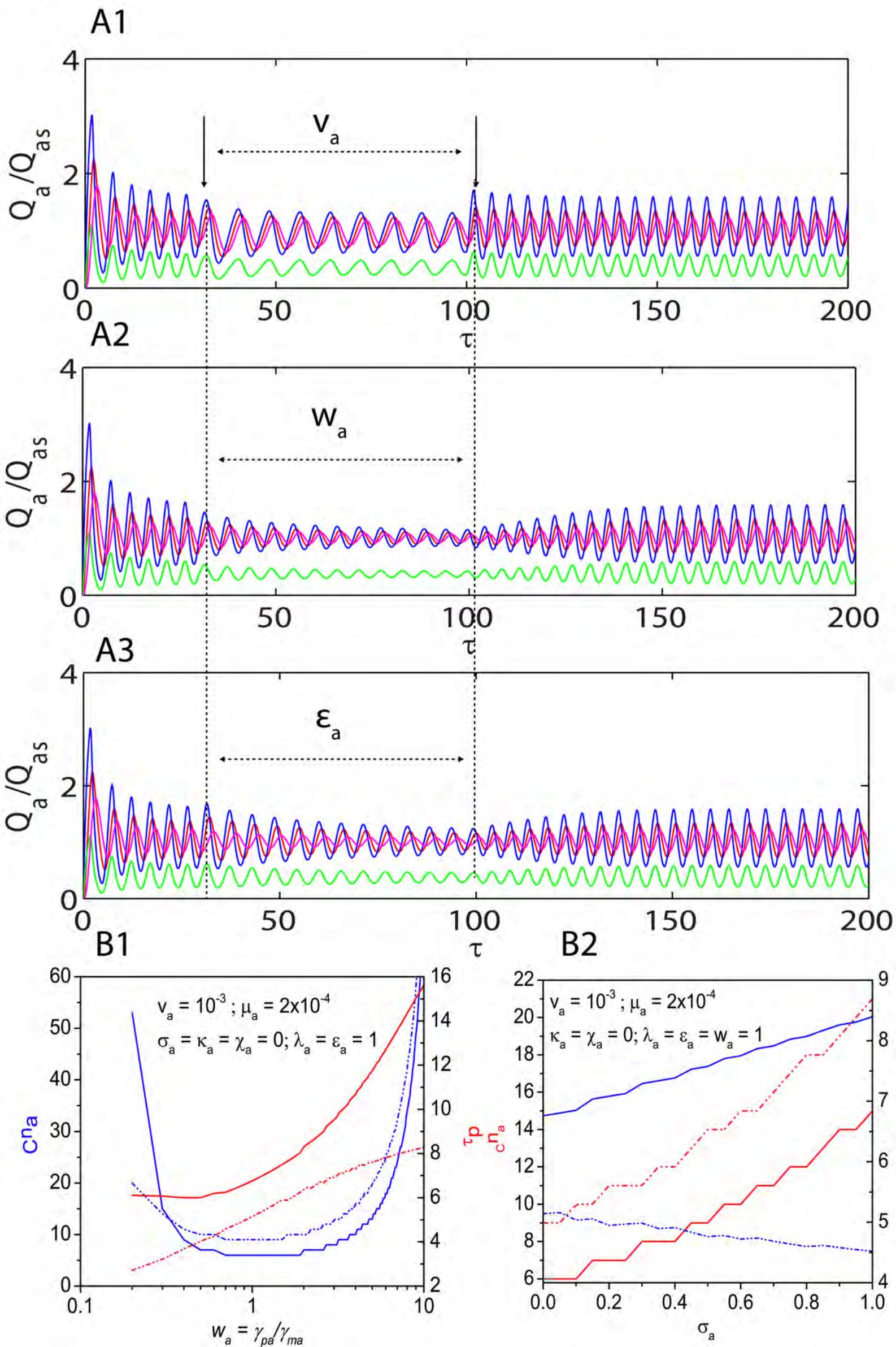

Figure 6

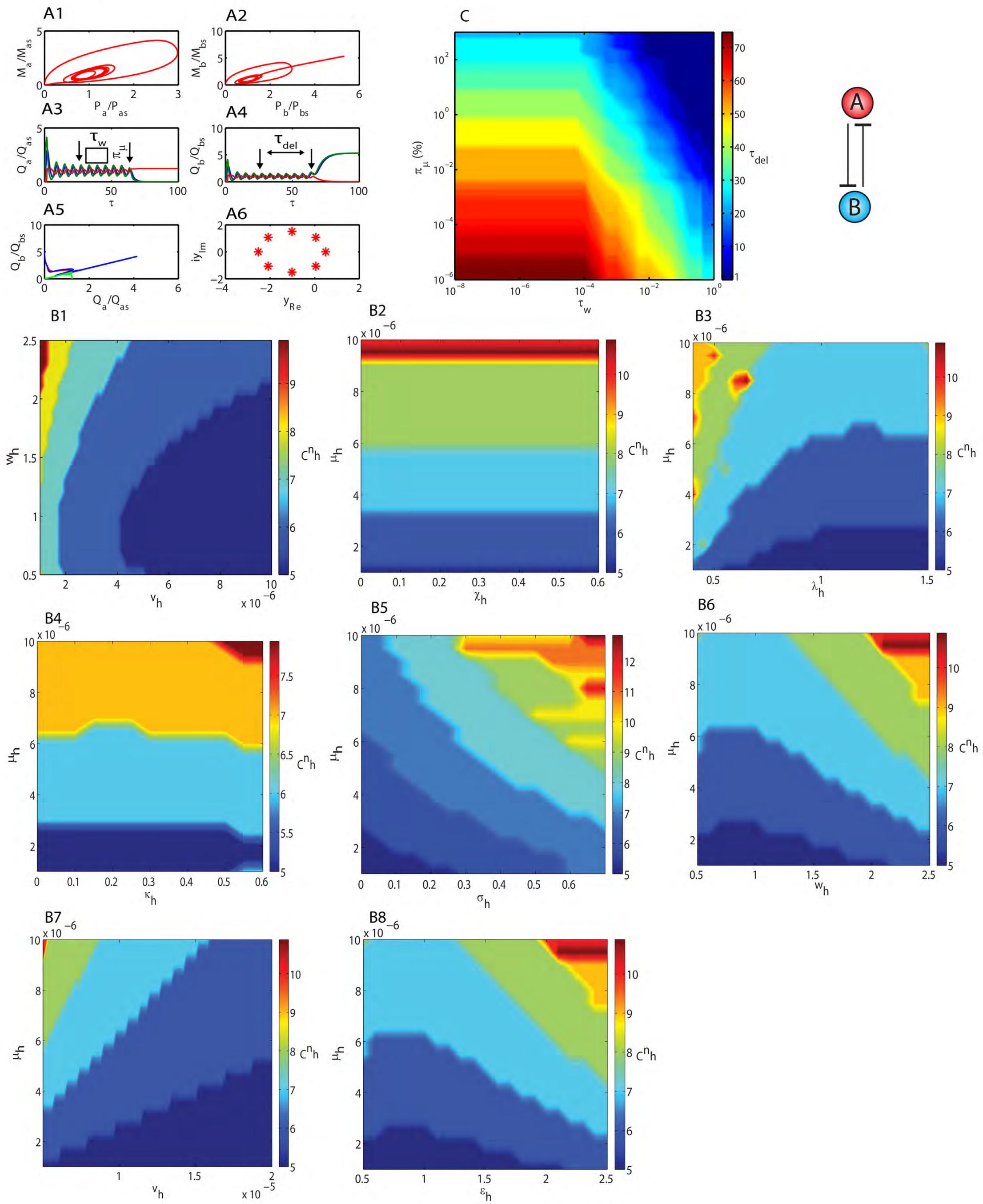

# Figure 7

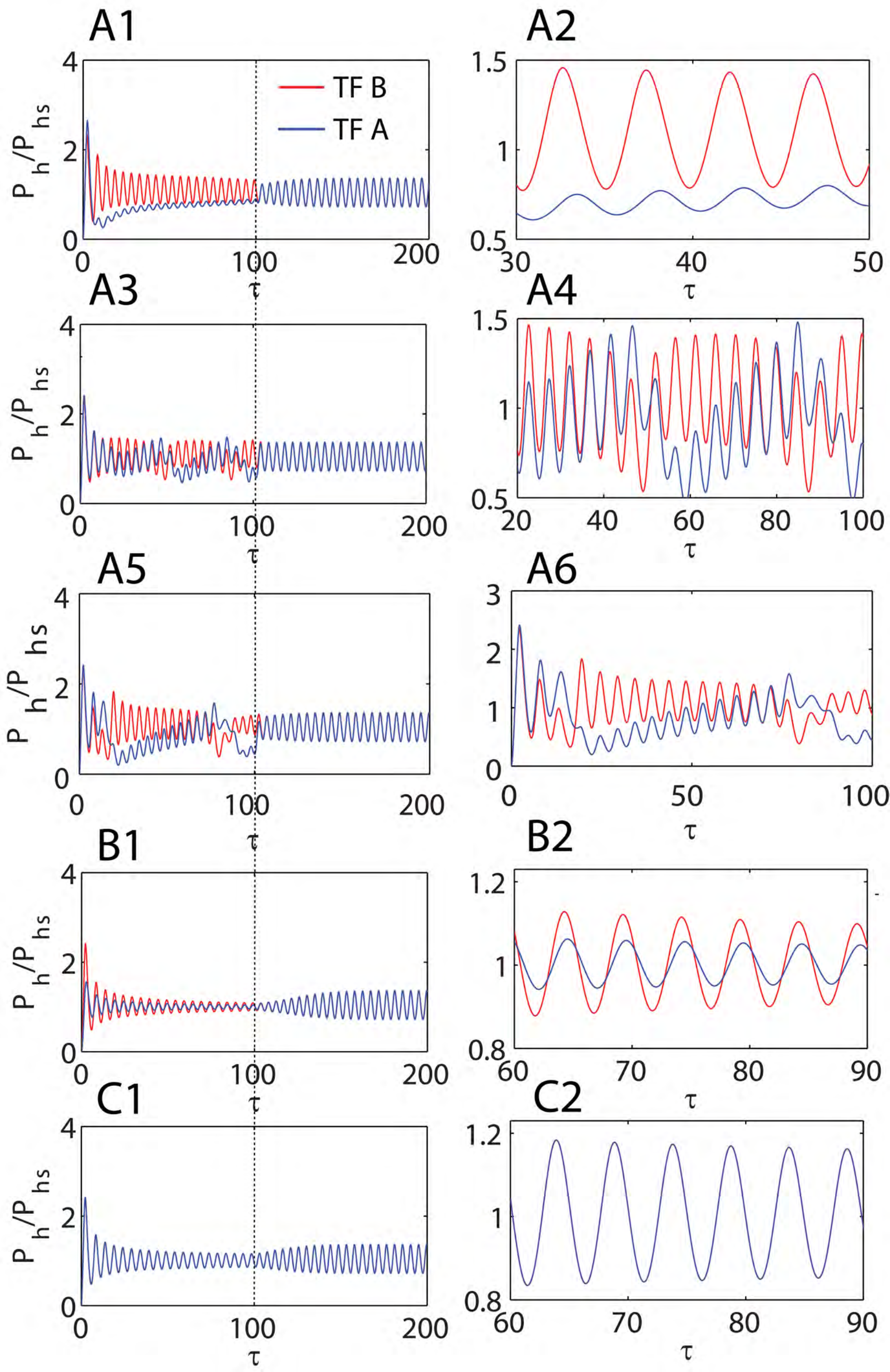

# Figure 8

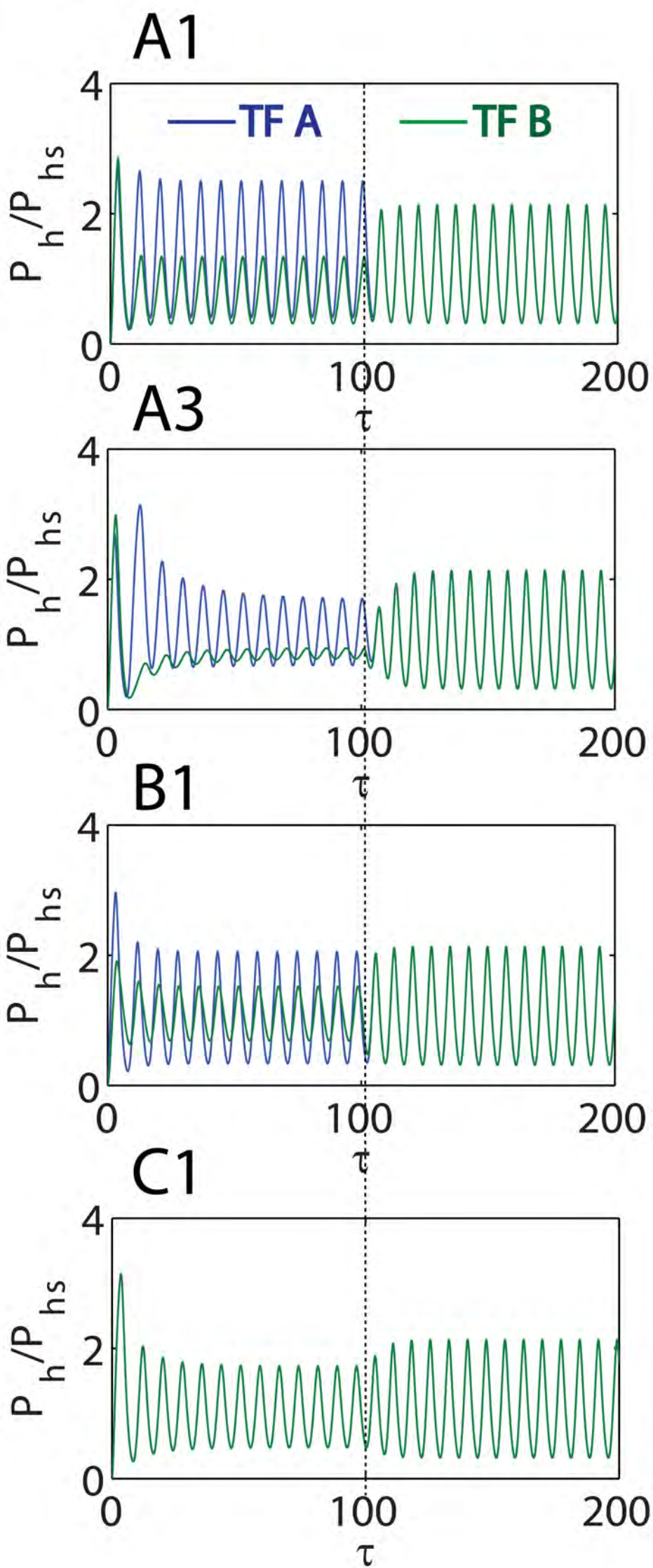
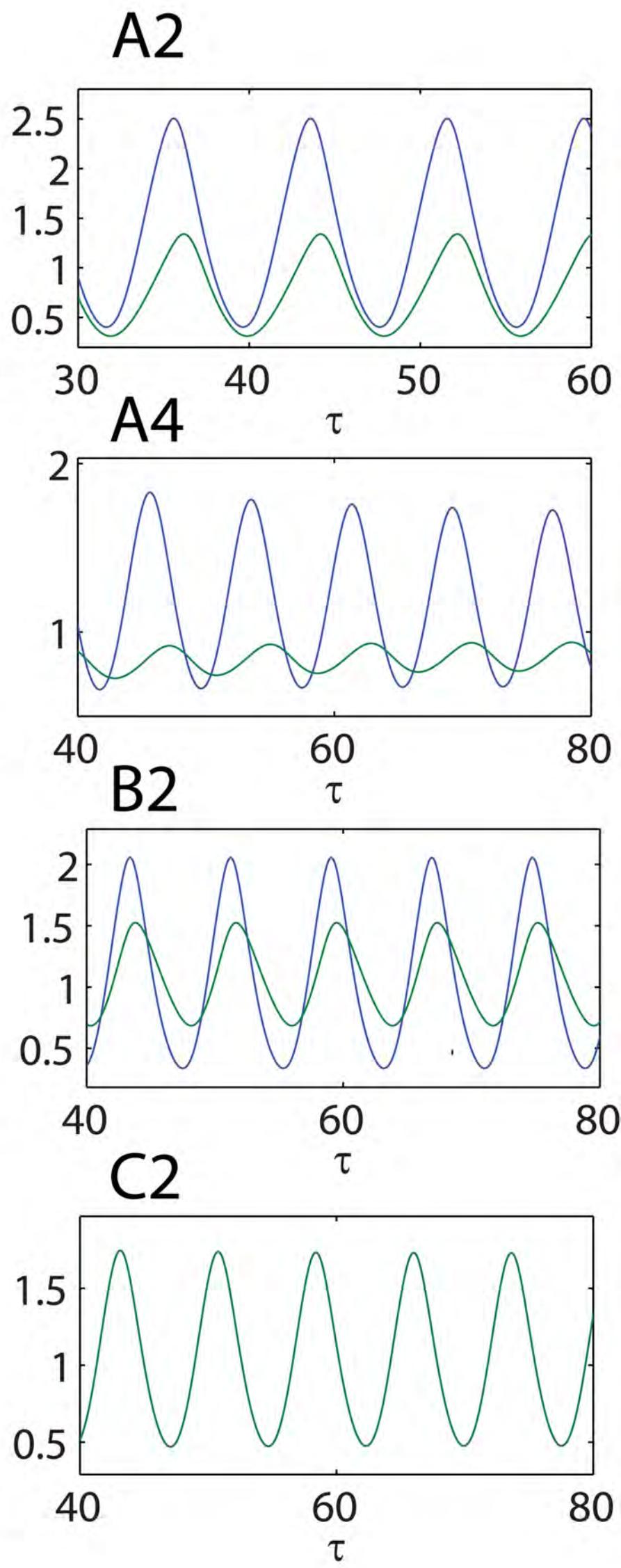

Figure 9

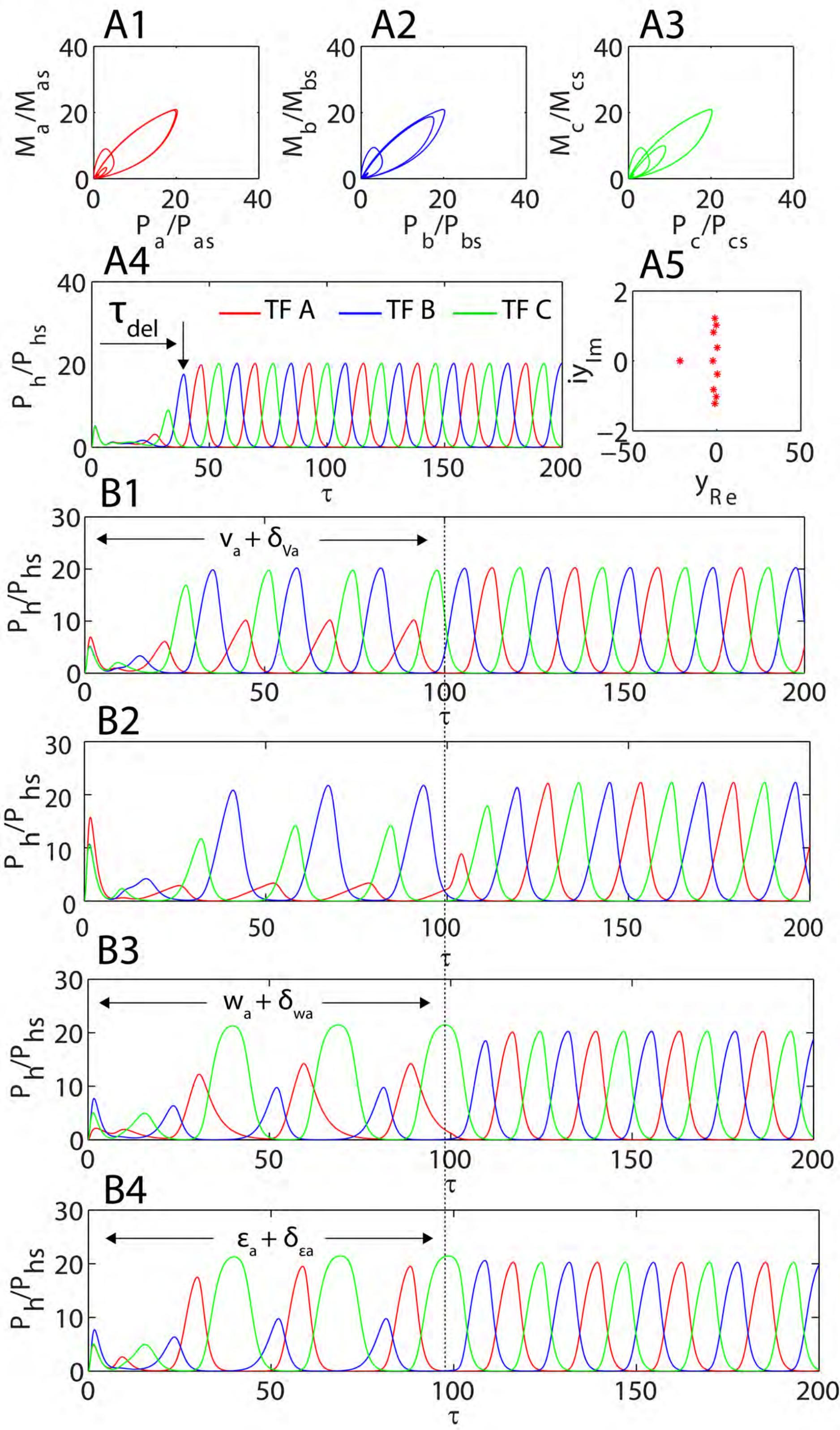

**Figure 10**

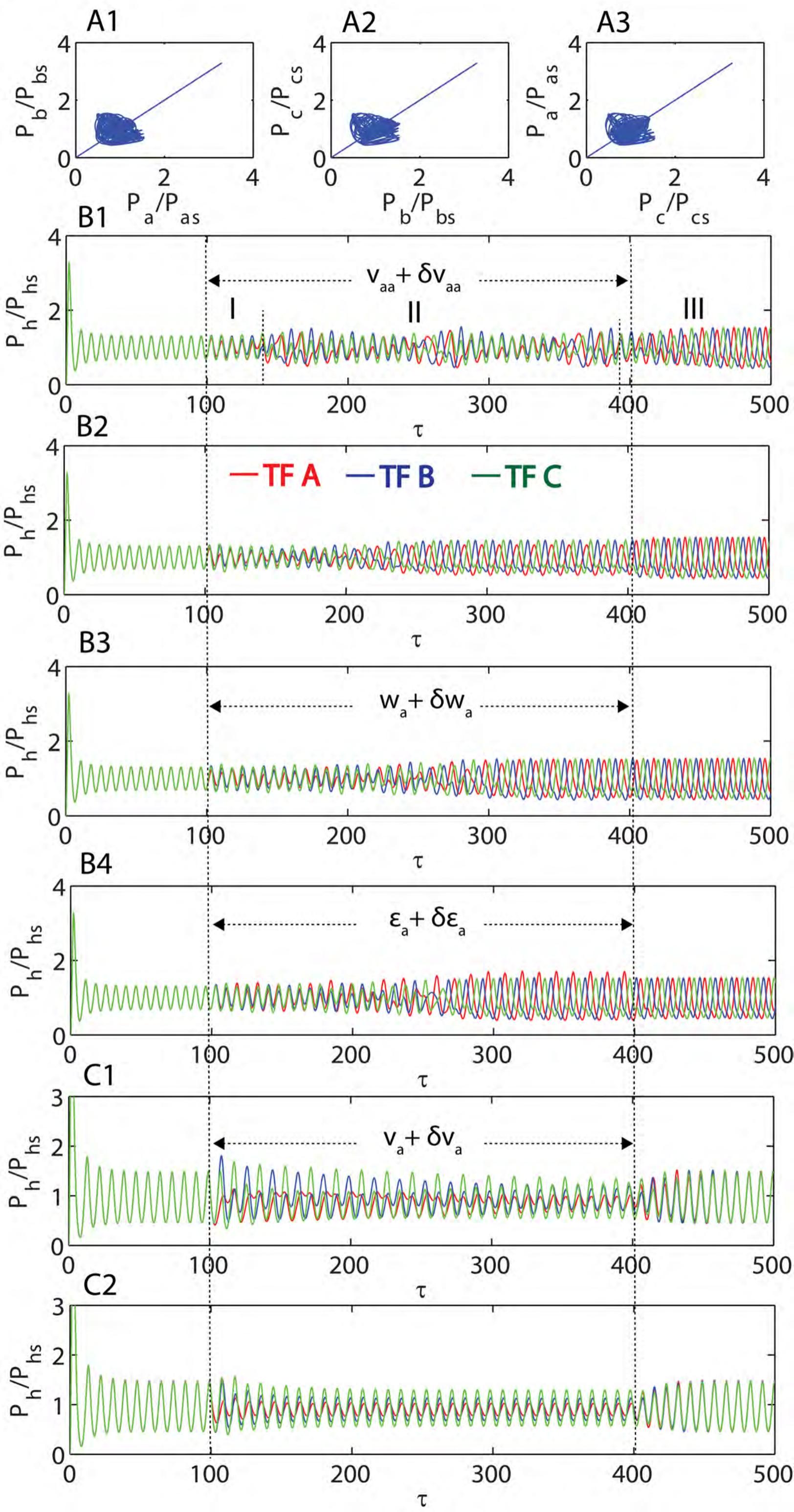

# Figure 11

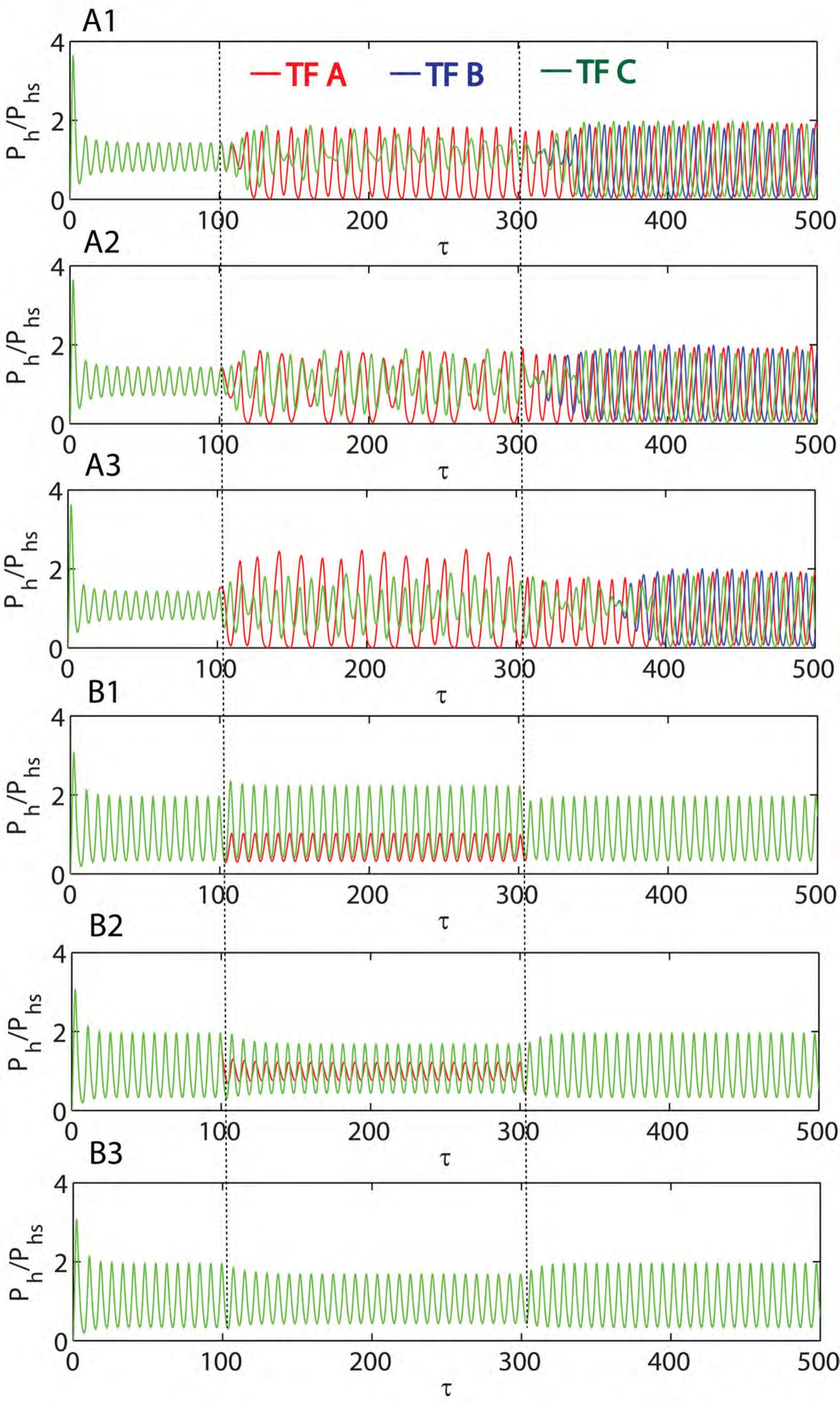

# Figure 12

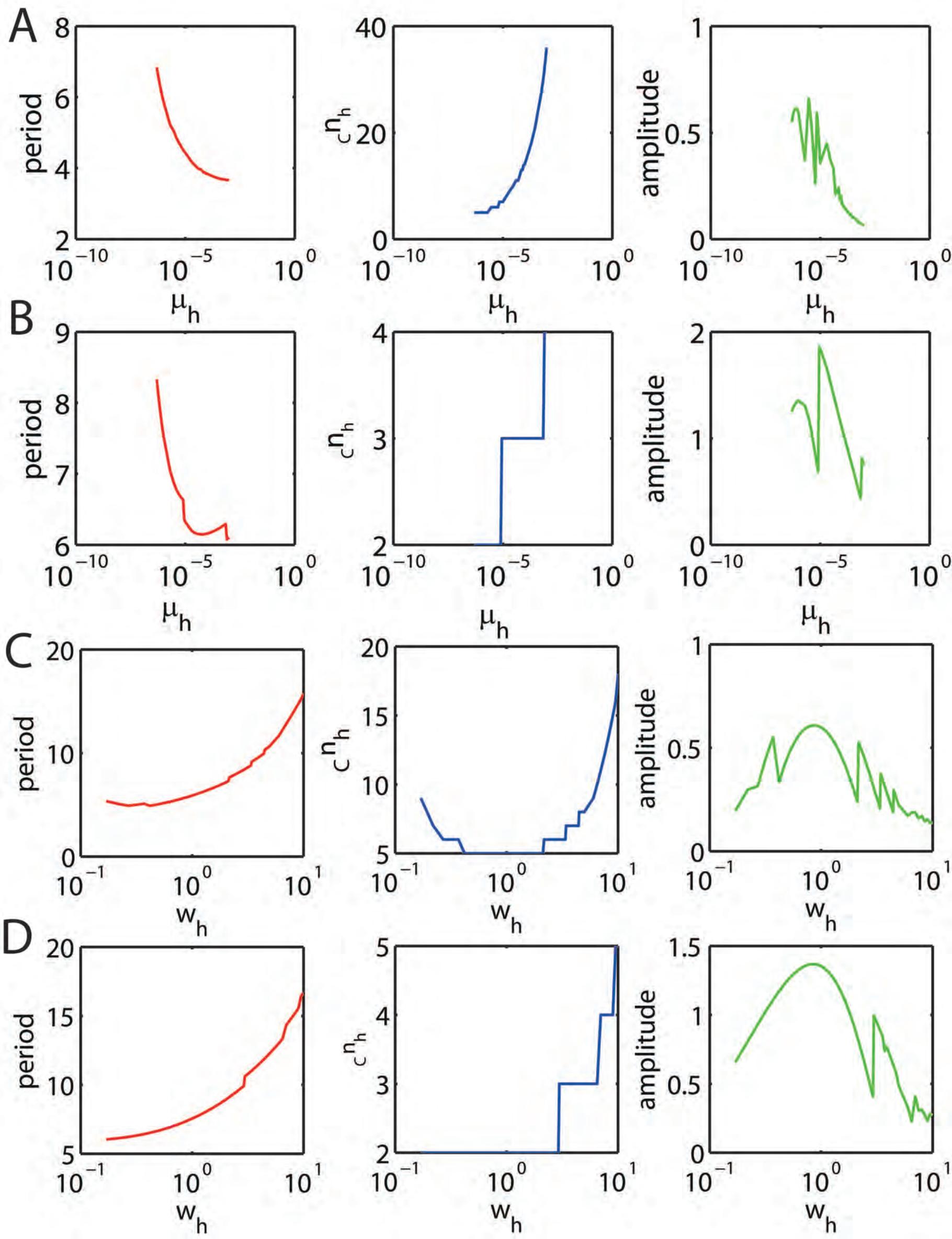

# Figure 13

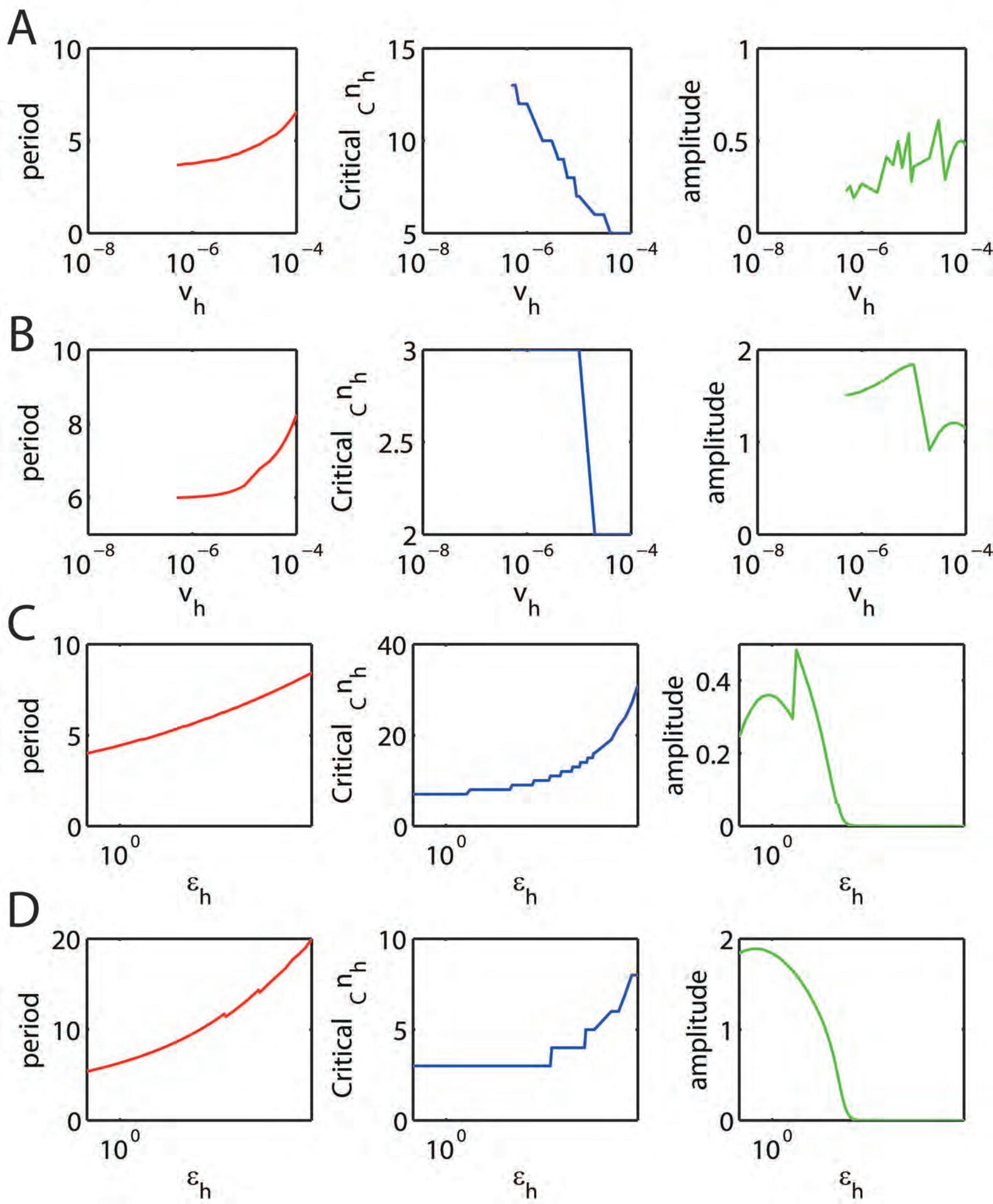